\documentclass[prc,reprint,twocolumn,amsmath,superscriptaddress]{revtex4-1}

\usepackage[latin1]{inputenc}
\usepackage{natbib}
\usepackage{amsmath}
\usepackage{amsfonts}
\usepackage{amssymb}
\usepackage{graphicx}
\usepackage{lineno}
\usepackage{amsmath}
\usepackage{multirow}
\usepackage{siunitx}
\usepackage[]{units}
\usepackage{braket}
\usepackage{rotating}
\usepackage{colortbl}
\usepackage{xcolor}
\usepackage{fullpage}
\usepackage{pbox}
\usepackage{url}
\usepackage{wasysym}
\usepackage{enumitem}
\usepackage{hyperref}
\usepackage{array,booktabs,tabularx}
\newcolumntype{C}{>{\centering\arraybackslash}X}
\usepackage[capitalize]{cleveref}

\crefname{figure}{fig.}{figs.}
\crefname{table}{table}{tables.}
\crefname{equation}{eqn.}{eqns.}
\crefname{section}{sec.}{secs.}

\begin{document}

\title{Experiment to Demonstrate Separation of Cherenkov and Scintillation Signals}

\newcommand{\ucb}{University of California, Berkeley, CA 94720-7300, USA}
\newcommand{\lbl}{Lawrence Berkeley National Laboratory, CA 94720-8153, USA}
\newcommand{\bnl}{Brookhaven National Laboratory, Upton, NY 11973-500, USA}
	
\affiliation{\ucb}
\affiliation{\lbl}
\affiliation{\bnl}

\author{J. ~Caravaca}\affiliation{\ucb}\affiliation{\lbl}
\author{F.B. ~Descamps}\affiliation{\ucb}\affiliation{\lbl}
\author{B.J. ~Land}\affiliation{\ucb}\affiliation{\lbl}
\author{J. ~Wallig}\affiliation{\lbl}
\author{M. ~Yeh}\affiliation{\bnl}
\author{G.D. ~Orebi Gann}\affiliation{\ucb}\affiliation{\lbl}

\begin{abstract}
The ability to 
separately identify the  Cherenkov and scintillation light components produced in scintillating mediums holds the potential for a major breakthrough in neutrino detection technology, allowing development of a large, low-threshold, directional detector with a broad physics program.  
The CHESS (CHErenkov / Scintillation Separation) experiment employs an innovative detector design with an array of small, fast photomultiplier tubes and state-of-the-art electronics to demonstrate the reconstruction of a Cherenkov ring in a scintillating medium based on photon hit time and detected photoelectron density. 
This paper describes the physical properties and calibration of CHESS along with first results.  The ability to reconstruct Cherenkov rings is demonstrated in a water target, and a time precision of $338\pm12$~ps FWHM is achieved.    Monte Carlo based predictions for the ring imaging sensitivity with a liquid scintillator target predict an efficiency for identifying Cherenkov hits of $94\pm1$\% and $81\pm1$\% in pure linear alkyl benzene (LAB) and LAB loaded with 2 g/L of PPO, respectively, with a scintillation contamination of $12\pm1$\% and $26\pm1$\%.

\end{abstract}

\maketitle

\section{Introduction}

Optical photon detection has been a common detection mechanism in neutrino experiments for many decades~\cite{imb,superk,sno,kamland,borexino,lsnd} and the technology is well developed. Experiments have historically been optimized to detect one of two types of optical radiation: Cherenkov~\cite{cherenkov} or scintillation~\cite{birks} light.  Directional Cherenkov light has been successfully used in both high energy and nuclear physics by large ultra-pure water experiments such as SuperKamiokande~\cite{superk} and SNO~\cite{sno}, while the more abundant, isotropic scintillation is more commonly used in low-energy detectors such as KamLAND~\cite{kamland} and Borexino~\cite{borexino}.  A brief summary of the advantages and limitations of each technique can be found in \Cref{tab:chervsscint}.

\bgroup
\def\arraystretch{1.6}
\begin{table}[th!]
{\scriptsize 
	\def\tabularxcolumn#1{m{#1}}
	\begin{tabularx}{\linewidth}{*{2}{C}}
	   Cherenkov & Scintillation \\
	\hline
	\hline
	  Directional & Isotropic\\
	\hline
	  Very well understood  & Strong dependence on material properties  \\
	\hline
	  Good shower/MIP separation & Reasonable particle ID  \\
	\hline
	  Minimum energy threshold for light production & No energy threshold for light production  \\
	\hline
	  Low light yield & High light yield results in lower detector threshold and improved energy resolution \\
	\hline
	  Occurs in all dielectric materials, some with very good optical properties & Scintillating materials tend to have substantially shorter attenuation lengths  \\
	\hline
	  Cost-effective & More expensive materials \\
	\end{tabularx}
}
\caption{Comparison of Cherenkov and scintillation light in the context of optical particle detection.}
\label{tab:chervsscint}
\end{table}
\egroup

The LSND experiment~\cite{lsnd} used a mineral-oil-based scintillator designed to allow detection of both scintillation and Cherenkov light, opening up  sensitivity to low-energy particles  below  Cherenkov threshold. The ratio of created photoelectrons for the two sources of light was approximately 5 to 1 (scintillation to Cherenkov) for a 45~MeV electron created in the detector. Photon timing and charge was used for particle identification and to reconstruct vertex and angle information. Particles emitting Cherenkov radiation produced significant fractions of prompt light and thus particle identification techniques relied on the fraction of prompt light as a tool for discrimination. \\
Separation of Cherenkov- and scintillation-light components on an event-by-event basis enables a single detector to exploit the advantages of each technique.  If this separation can be achieved in a pure liquid scintillator (LS) detector it would substantially improve the sensitivity of low-energy physics programs by adding directional reconstruction capability to a low-threshold detector. Searches for neutrinoless double beta decay (NLDBD) would benefit from rejection of the directional solar neutrino signal, the dominant background in experiments such as SNO+~\cite{snop}.  Precision solar neutrino measurements would benefit from  separation of the directional signal from isotropic radioactive background events. 
However, the high scintillation light yield and fast timing of commonly used LS such as LAB (linear alkyl benzene) makes this separation extremely challenging.  In addition, liquid scintillator detectors are limited to kiloton scales by the attenuation of light in the scintillator material and also by cost.

The recent development of Water-based Liquid Scintillator (WbLS)~\cite{wbls} allows both the scintillation light yield and timing profile of the target material to be tuned, thus increasing sensitivity to the Cherenkov component.  The admixture of water increases the attenuation length, thus improving light collection.  For the first time one can thus envisage a large-scale, low-threshold detector with directional sensitivity.  
The concept for a large monolithic WbLS detector capable of a very broad program of physics, such as the \textsc{Theia} experiment, is presented in~\cite{asdc} and ~\cite{theia}.  The unique capabilities of a large-scale WbLS detector enable both a broad low-energy program, including a sensitive NLDBD search, solar neutrino studies, supernova neutrinos, and diffuse supernova neutrinos (DSNB), as well as sensitivity to nucleon decay and, if sited in a high-energy neutrino beam, to long-baseline physics (neutrino mass hierarchy and CP violation).  The high-energy program benefits from the potential to image Cherenkov rings, thus improving particle ID, and the ability to detect particles below Cherenkov threshold, which improves background rejection.

The goals of the CHErenkov / Scintillation Separation (CHESS) experiment are several: to demonstrate Cherenkov / scintillation separation in a pure LS target by deploying ultra-fast-timing photosensors, and to study how the separation improves in WbLS and quantify the efficiency as a function of the scintillator fraction.  These studies will allow optimization of the \textsc{Theia} detector configuration (WbLS target cocktail, photosensor requirements, detector scale) in order to maximize the physics reach across a broad program. 

\begin{figure}
	\centering
	\includegraphics[width=.45\textwidth]{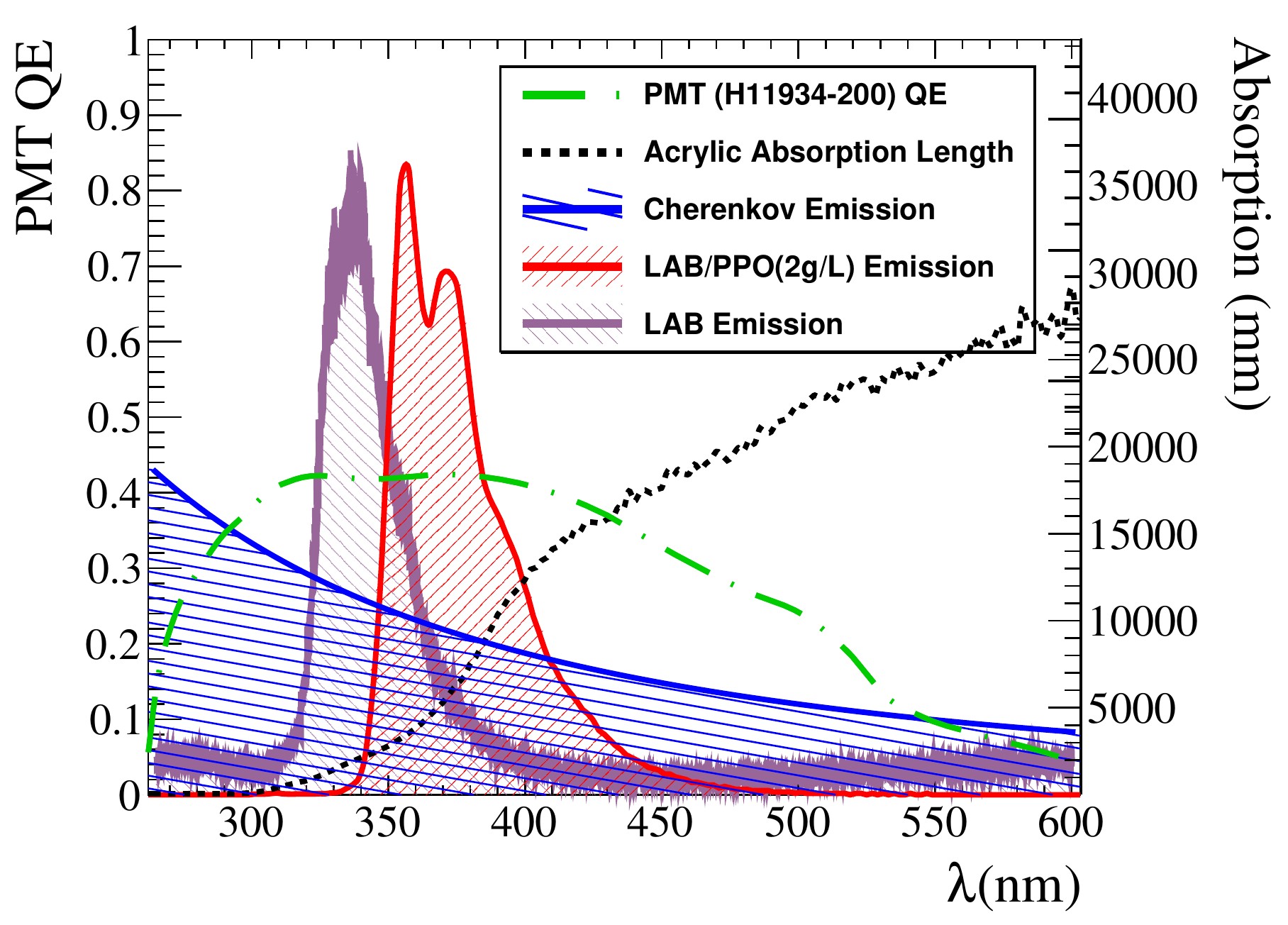}
	\caption{\label{fig:optics} PMT H11934 quantum efficiency~\cite{h11934} and UVT acrylic absorption length compared to the Cherenkov and scintillation emission spectra for pure LAB~\cite{lab_emission} and LAB loaded with 2~g/L PPO~\cite{snop_private}. The normalization of the emission spectra are shown in arbitrary units.}
\end{figure}

Separation of scintillation and Cherenkov components can occur in three domains:
\begin{itemize}
    \item Wavelength:  Cherenkov and scintillation light have  different emission spectra (\Cref{fig:optics}). The Cherenkov signal could be enhanced by selecting photosensors  with improved sensitivity in the high wavelength region above the scintillation emission cutoff.
	\item Photon density: depending on the ratio of Cherenkov to scintillation light yield, it may be possible to distinguish the  Cherenkov ring structure on top of the background of isotropic scintillation light. 
	\item Time separation: scintillation light, originating from molecular de-excitation, is typically delayed from the fast Cherenkov light pulse by 1 to 10s of ns~\cite{cherenkov, birks}. By combining fast photon detectors and electronics it should be possible to achieve some signal separation using photon detection times.  
\end{itemize}

Each option has the potential to enhance signal separation, but each comes with a cost, such as: reduced light yield, reduced light collection efficiency, high-cost photosensors, or limitations on detector size.  The challenge for time-based separation is the timing precision of  available photosensors. Large photomultiplier Tubes (PMTs) typically have worse than ns  time precision, making Cherenkov light identification in a scintillation medium extremely difficult. However, newly developed small PMTs~\cite{h11934} can achieve  timing precisions of $\lesssim$ 300~ps and new micro-channel plate (MCP) photosensor technology~\cite{mcp, lappd, lappd2} can achieve  $\lesssim$ 100~ps. In addition, fast and precise commercial electronics are available to digitize these signals without significantly degrading the time resolution. This makes  separation in the time domain a  theoretical possibility.

This paper describes the CHESS detector and its sensitivity to Cherenkov / scintillation separation in  the time domain.  Section~\ref{s:desc} describes the detector itself.  Section~\ref{sec:simulation} describes the detailed Monte Carlo simulation used to model the detector. 
Section~\ref{s:analysis} describes the general approach to data analysis.  Section~\ref{sec:calibration} describes detector calibrations.  Section~\ref{s:event} describes event selection and techniques for rejection of instrumental and physics background events.  Section~\ref{s:prospects} describes the predicted sensitivity in a pure LS target: for both pure LAB and LAB loaded with 2~g/L of PPO (hereafter referred to as LAB/PPO); and Section~\ref{s:conc} concludes the paper.

\section{The CHESS Detector}\label{s:desc}

\begin{figure}
\centering
\includegraphics[width=\columnwidth]{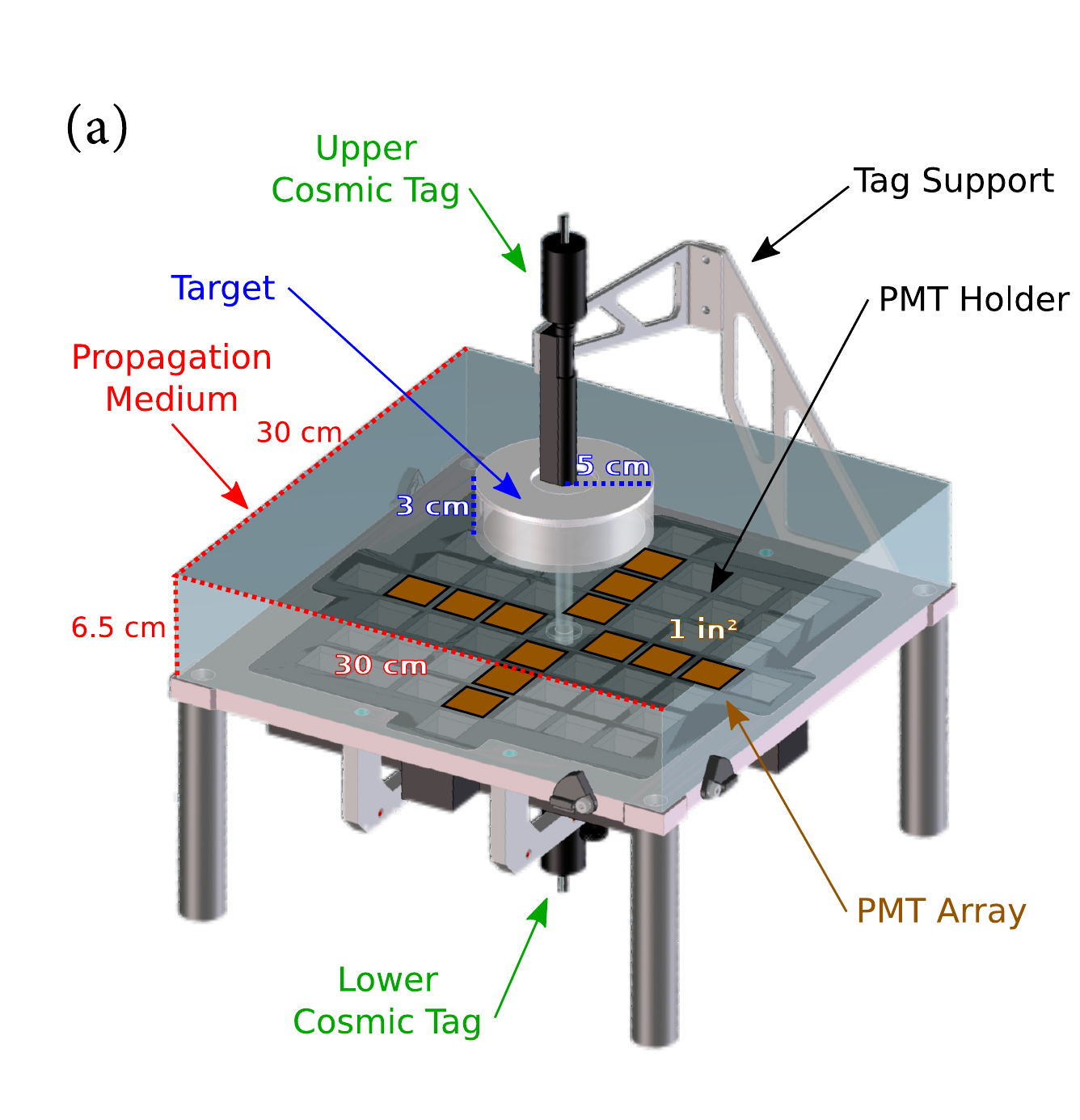}
\includegraphics[width=0.89\columnwidth]{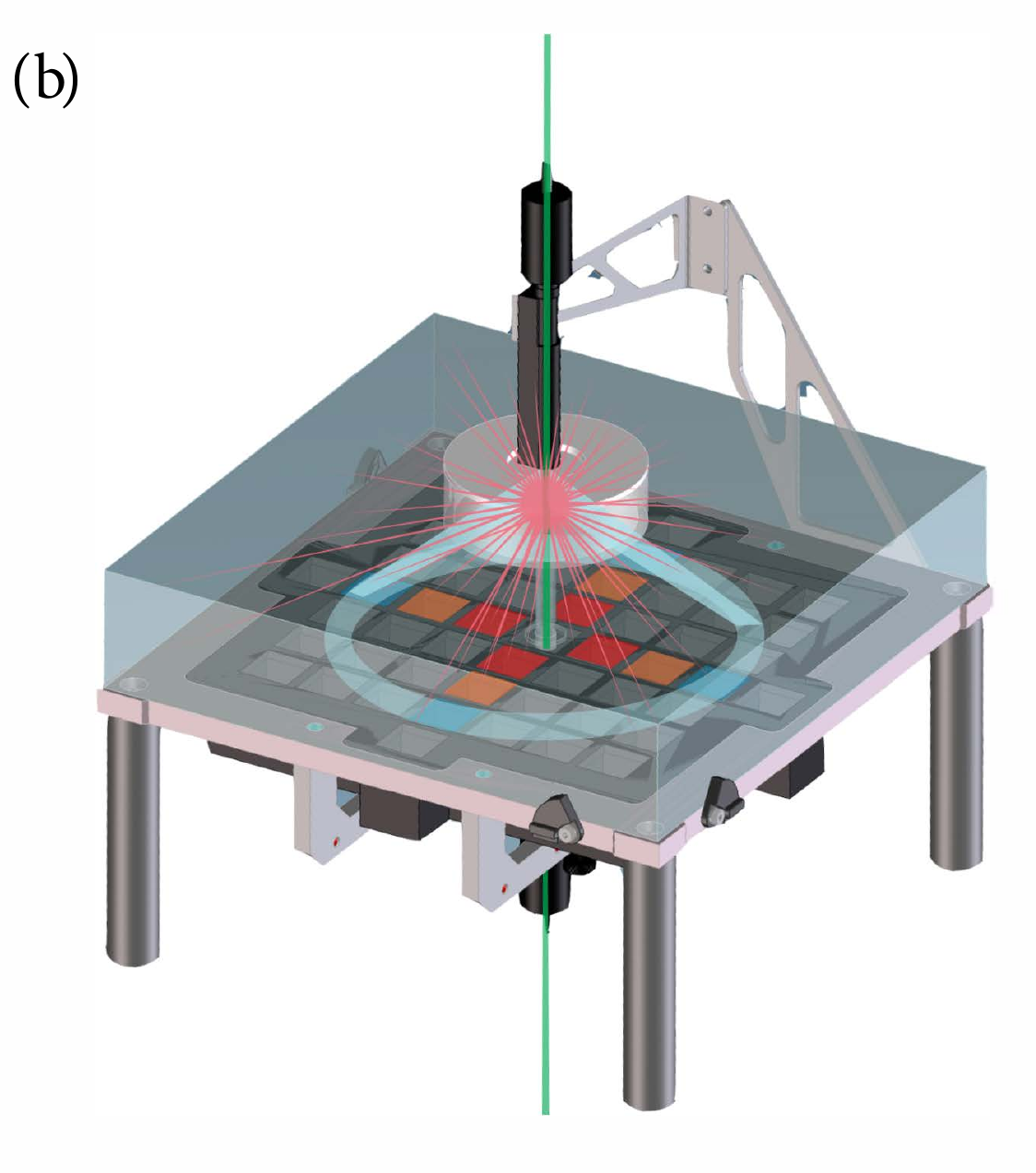}
\caption{The CHESS apparatus.  (a) Detailed schematic view with dimensions and (b) demonstration of ring-imaging concept. The PMT array is designed to hold up to 53 PMTs; the dozen slots occupied for this study are color coded by radius: red and orange for those hit primarily by scintillation photons, and blue for those in the expected Cherenkov ring for LAB and LAB/PPO. Due to the lower refractive index, the ring from a water target is detected in the middle (orange) PMTs. }
\label{fig:timing-setup}
\end{figure}

The primary goal of the CHESS experiment is to demonstrate Cherenkov / scintillation  separation by Cherenkov ring imaging in both charge (detected photoelectron density) and time, for different scintillating liquids.  
A schematic of CHESS is shown in \Cref{fig:timing-setup}.  An acrylic target vessel is viewed by an array of small, fast PMTs. 
The setup is designed to detect either cosmic muons or events from deployed radioactive sources.  The primary ring imaging measurement is performed using through-going muons.  Vertical-going events are selected via a 1-cm diameter coincidence tag, ensuring a population of events with known orientation and thus a known expectation for the position of the Cherenkov ring. 
The muons produce Cherenkov and scintillation light in the target material, which is detected on the PMT array. The apparatus has been optimized using a full Monte Carlo simulation (Sec.~\ref{sec:simulation}), complete with optical ray tracing, such that direct Cherenkov light from vertical muons falls on a distinct set of PMTs, forming a clear ring in the PMT array.  This yields two distinct groups of PMTs by construction: those with pure scintillation hits and those with both scintillation and Cherenkov hits. The earliest hits on each PMT can thus be identified as being caused by either Cherenkov or scintillation photons, and demonstrate the time separation between these two signals to high precision. 
A  measurement of the clarity of the Cherenkov ring imaged in charge on top of the isotropic scintillation light background is also possible.

\subsection{Target Vessel}
The target vessel  consists of a cylinder 5~cm in radius and 3~cm in height, constructed from ultraviolet-transmitting (UVT) acrylic. Three flat faces on the outer surface of the cylinder, each 3~cm in diameter, provide surfaces for attaching a radioactive button source or optical coupling of a PMT as a source tag. A lid made of the same UVT acrylic encloses the target material in an air-tight environment using an FFKM o-ring~\cite{cog-oring}. 

Distinct but similarly designed target vessels are used for water, pure LS and WbLS target materials in order to minimize the risk of cross contamination.

\subsection{PMT Array}

\label{pmtarray}
In the initial phase of CHESS, one dozen Hamamatsu H11934-200 PMTs~\cite{h11934} were deployed in a cross shape beneath the target, providing three radial rings of four PMTs each for detection of Cherenkov and scintillation light.  An additional twelve similar PMTs are available for deployment in a future upgrade.  The H11934-200 PMT is a small 1-inch cubic PMT with superb quantum efficiency (QE) peaked at $42\%$ (\Cref{fig:optics}) and excellent transit time spread (TTS) of 300~ps (FWHM for single photoelectrons).

The PMTs are held in an array that consists of a 7x7 grid, with four additional slots at each compass point, as shown in~\Cref{fig:timing-setup}.  The holder is 3D printed from black ABS plastic.  The initial deployment positions for the twelve PMTs used in phase I of CHESS are shown in \Cref{fig:timing-setup}.
The unfilled locations on the grid can be populated with additional PMTs for future expansion. 

\subsection{Optical Propagation}
The target vessel is optically coupled to a propagation medium made of UVT acrylic (30~cm$\times$30~cm$\times$6.5~cm). This coupling eliminates an acrylic/air boundary which would otherwise block all Cherenkov light due to total internal reflection. The propagation medium acts as a light guide to allow photons to propagate from the bottom of the target vessel towards the PMT array. The PMTs in the array are similarly optically coupled to the bottom of the propagation medium. The propagation medium and target vessel have been polished for a better optical coupling and a more accurate simulation of refracted light at the boundaries. 

A vertical cylindrical hole 1~cm in diameter is machined through the center of the propagation medium aligned with the center of the target vessel and cosmic tags in order to allow  vertical-going  muons to interact with the target material, but pass through the propagation medium without producing additional Cherenkov light in the acrylic block that would contaminate the measurement.

Throughout the setup, EJ-550 optical grease~\cite{ej550} is used for optical coupling of components.

\subsection{Cosmic Muon Tags}

Two custom-made cylindrical scintillator tags are positioned above and below the target vessel (\Cref{fig:timing-setup}) in order to trigger on vertical cosmic muons.  An aluminum arm maintains the alignment of the upper cosmic muon tag while the bottom is fixed in the PMT array (described in \Cref{pmtarray}). Each tag consists of a cylindrical 1-cm diameter Hamamatsu PMT~\cite{h3164} optically coupled to EJ-200 plastic scintillator~\cite{ej200} shaped into a cylinder of 1-cm diameter and 5-cm height. The scintillator is coated with white paint to reflect light into the tag PMT, and then coated with matte black paint (except for the end that is coupled to the PMT) to shield the remainder of the setup from light contamination.  This assembly is held in a custom-built mount.

The small size of the tags results in a low angular acceptance (6$\textdegree$ from vertical), ensuring a population of events with known orientation and thus a known expectation for the position of the Cherenkov ring.  
Given the typical muon flux at the Earth surface ($1~\mbox{cm}^{-2}\mbox{min}^{-1}$) and the angular acceptance of the tags, a coincidence rate of $\sim4~\mu / \mbox{day}$ is predicted. 

\subsection{Veto Panels}\label{s:veto}

\begin{figure}
\centering
\includegraphics[width=0.9\columnwidth]{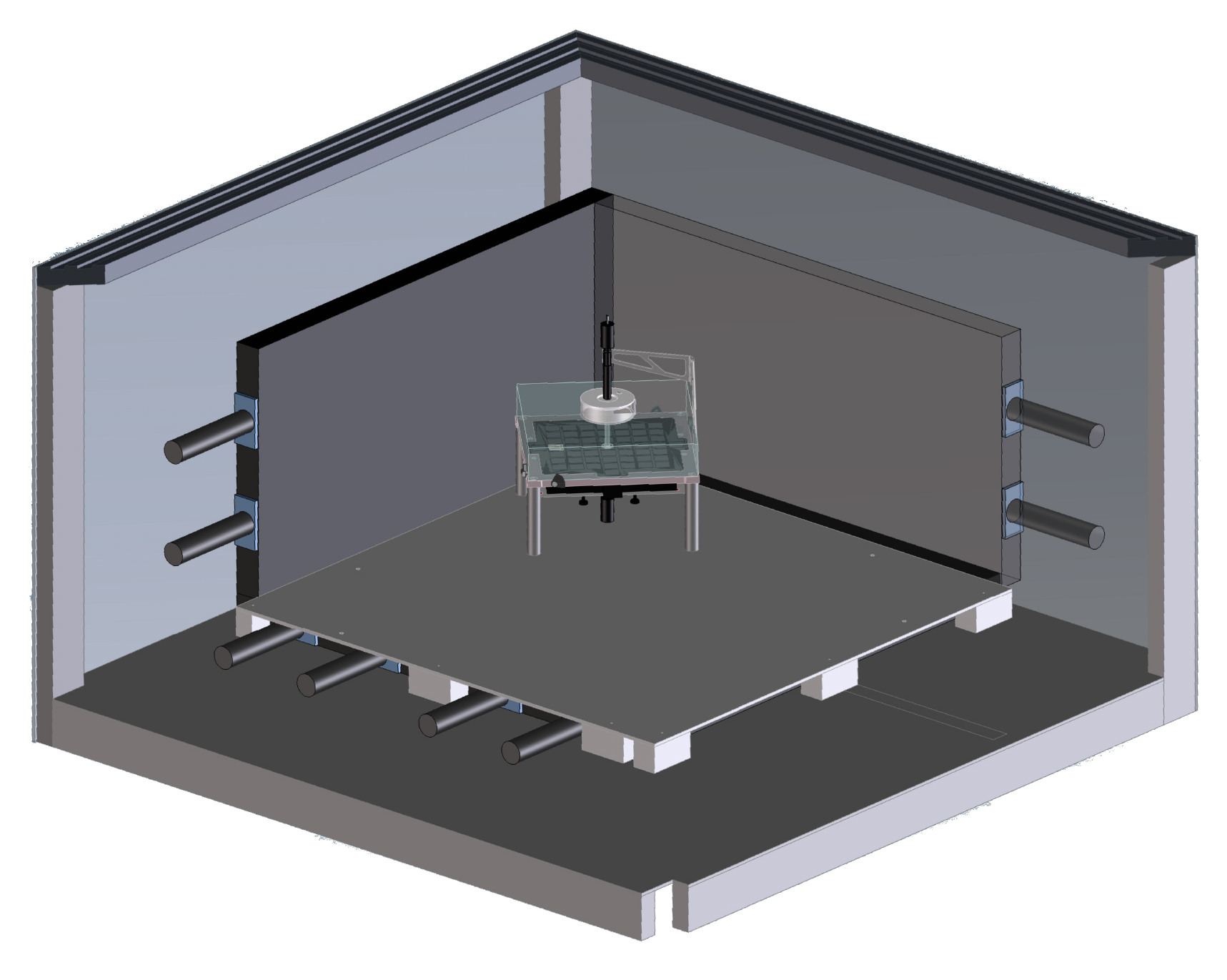}
\caption{Layout of veto panels around the CHESS apparatus. }
\label{f:veto}
\end{figure}

The apparatus is surrounded by four scintillator panels (50~cm$\times$100~cm$\times$5.3~cm), as shown in Fig.~\ref{f:veto}, fabricated from EJ-200 plastic scintillator~\cite{ej200} (two on the floor and two on the sides in a corner distribution) providing effective $4\pi$ coverage. 
Each panel is instrumented with a PMT~\cite{9102ksb} which is read out for each event and used for an offline veto. The scintillator panels are used to veto cosmic events during calibration source deployment, and to reject cosmic shower events and coincidence of multiple cosmic muons in the ring imaging analysis. 

\subsection{Shielding}
CHESS is enclosed in a light-tight darkbox of about 1.5~m$\times$1.5~m$\times$1~m wrapped in FINEMET\textsuperscript{\textregistered} sheets~\cite{finemet} for magnetic field isolation and painted matte black on the inside to minimize reflections.

\subsection{Radioactive Source Deployment}\label{s:source}
A $^{90}$Sr button source is used to calibrate various aspects of the detector response (Sec.~\ref{sec:calibration}).  The source consists of a 1-inch diameter acrylic disk with a thickness of 0.125 inches (\cite{buttonsource}). A 0.25-inch diameter cylindrical well in the center of the button contains a deposit of 0.1$\mu Ci$ of $^{90}$Sr combined with a resin. The well containing the radioactive material is encapsulated with a 0.02-inch thick acrylic window. This source is attached to one face of the target vessel with the window facing the target material.

To trigger on source events a  1-inch cubic H11934-200  PMT~\cite{h11934} is optically coupled to another of the flat faces on the target vessel. This is the same type of PMT used in the PMT array and was chosen for its high QE and low TTS. 

\subsection{The DAQ System \label{sec:daq}}

\begin{figure}
\includegraphics[width=\columnwidth]{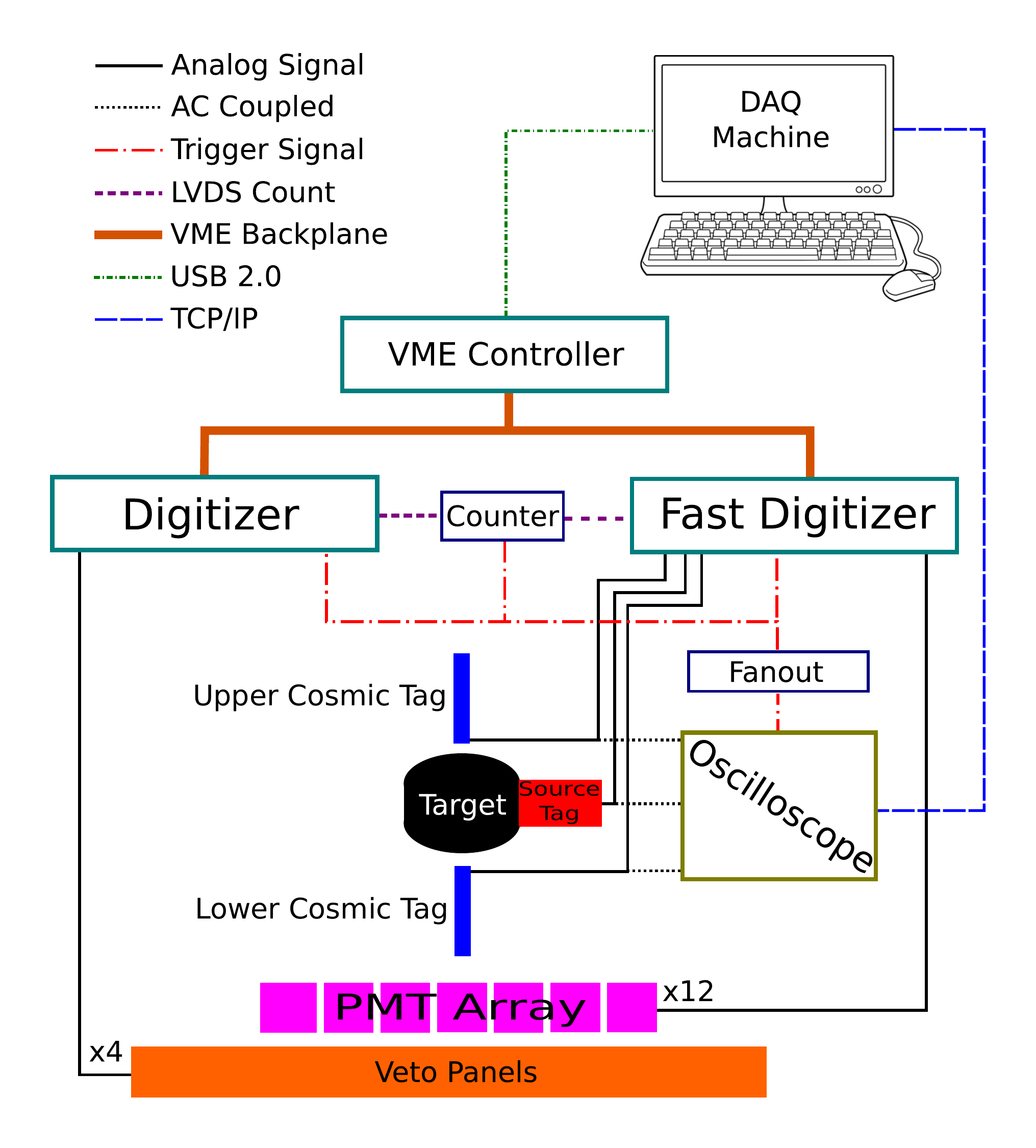}
\caption{A schematic diagram of how the DAQ and hardware are connected together with signal and data paths labeled. Omitted are the high voltage supplies, which are connected to the VME backplane, and all PMTs.}
\label{fig:daq}
\end{figure}

A schematic diagram of the data acquisition (DAQ) hardware is shown in \Cref{fig:daq}  (high voltage omitted) and described in detail in the following sections. 
The DAQ software~\cite{wblsdaq} utilizes the CAEN VME library~\cite{caen-vme} to configure and read out the digitizers and high voltage supplies, and outputs HDF5~\cite{hdf5} formatted files containing raw digitized data and metadata for each triggered event.

\subsubsection{Readout Electronics and High Voltage}
Three six-channel high-voltage power supplies (CAEN V6533~\cite{v6533}) power the PMTs. 
The PMT output signals are connected to two CAEN digitizers: a high-precision V1730~\cite{v1730} digitizes the veto panels and source tag PMT signals, while a fast V1742~\cite{v1742} based on the DRS4~\cite{drs4} chip digitizes the PMT array and cosmic tag signals.  The hardware is housed in a VME crate, and a CAEN V1718~\cite{v1718} VME to USB bridge is used for communications.  

The V1742 card is capable of sub-$100$~ps resolution, which exceeds the TTS of the PMT array and is therefore not a limiting factor in the time precision.  However, on-board buffer size limits the acquisition to a maximum of 1024 samples or 200~ns. This shallow buffer necessitated a low-latency triggering scheme in order to contain the pertinent data in the available event window. The V1730 digitizer is deadtimeless, however the V1742 introduces dead time on the order of 100~$\mu$s.  This is neither a limitation for measurements with cosmic particles, where the trigger rate is approximately $0.2$~Hz, nor for measurements with radioactive source data where the trigger rate is approximately $30$~Hz.

\subsubsection{Trigger}
\label{sec:triggering}

A LeCroy 606Zi~\cite{lecroy606zi} oscilloscope is used to produce low-latency trigger signals for the setup with programmable coincidence logic. 
Depending on the operating mode one of the following three trigger configurations are used:
\begin{description}[font=\normalfont\textit]
\item [Bottom-only Trigger] A threshold condition is applied to the lower cosmic tag only. 
This is the normal operating mode for cosmic data, with the coincidence requirement applied offline.  
A rate of 4 $\mu$/day is expected after the coincidence trigger.
\item [Or Trigger] A threshold condition is applied to both of the cosmic tags, and the logical OR of the two is allowed to generate a trigger. 
This mode is used to acquire unbiased charge distributions for each cosmic tag simultaneously.
\item [Source Trigger] A threshold condition is applied to the source tag. 
This is the configuration used during radioactive source deployment.
\end{description}
The oscilloscope trigger signal is fanned out to the external trigger input on the V1730 and the low-latency trigger inputs on the V1742.
The V1742 consists of four DRS4 analog sampling chips recording eight channels each.
As these four chips operate on independent high-speed sampling clocks, the trigger signal is also digitized by each chip so that fine time offsets between channels on different chips can be reliably calculated offline.

\section{Monte Carlo Simulation}
\label{sec:simulation}

The entire setup is simulated in the RAT-PAC suite~\cite{ratpac}. Primary particles are produced (\Cref{sec:primarygen}) and tracked in a complete geometry implemented in GEANT4~\cite{geant4}. When Cherenkov and scintillation processes take place (\Cref{sec:photon_prod}), individual photons are tracked from the production point to the PMTs (\Cref{sec:optics}). Once the photon reaches the PMT, a custom model decides whether to produce a photoelectron (PE) based on the photon properties and PMT features (\Cref{sec:pmt_model}). If a PE is produced, a custom DAQ simulation (\Cref{sec:daq_sim}) produces a pulse per PE and the total waveform is digitized and stored for posterior analysis. Further details on each of the points are given in the following sections.

\subsection{Cosmic Muon Generator}
\label{sec:primarygen}

Muons and anti-muons are generated at $50\%$ ratio following the semi-empirical modified Gaisser distribution~\cite{gaisser-mod}. The hadronic component at surface is expected to be sub-dominant by a factor of approximately 50, resulting in a predicted 3 events within a month's worth of data, thus this component is not considered.  Muons are constrained to pass through the bottom tag volume.
There exists the possibility that high energy electrons produced by muon ionization trigger the bottom cosmic tag. This is reproduced in simulation by including a complete model of the holder material and geometry for both top and bottom tags.

\subsection{Photon Production \label{sec:photon_prod}}

Cherenkov production is simulated by GEANT4 by the standard class G4Cerenkov. The typical Cherenkov emission spectrum is shown in \Cref{fig:optics}. Cherenkov photons emitted with a wavelength below a certain value are immediately reabsorbed by the medium and hence are not propagated in the simulation. Scintillation emission is handled by a minimally modified version of the GLG4Sim model~\cite{glg4sim}. A charged particle passing through a medium deposits an energy $E$ due to ionization. The total number of scintillation photons generated is not expected to behave linearly with $E$ due to quenching effects. This is taken into account by using Birk's law~\cite{birks} which states that the deposited energy after quenching, $E_{q}$, is:
\begin{eqnarray}
	\dfrac{dE_{q}}{dx} = \dfrac{dE/dx}{1+ k_BdE/dx},
    \label{eq:birk}
\end{eqnarray}
where $k_B$ is Birk's constant. The total number of  photons produced, $N_{\gamma}$, is the direct product of $E_{q}$ and the light yield of the material in question in photons/MeV. Ex-situ measurements for the scintillation emission spectrum, light yield and time profile for LAB and LAB/PPO are included in the simulation. $N_{\gamma}$ photons are drawn randomly from the emission spectrum of the scintillator under investigation. The spectra for LAB~\cite{lab_emission} and LAB/PPO~\cite{snop_private} are shown in \Cref{fig:optics}. 

The short term components of the timing profile $\rho(t)$ for the scintillation process is well described by a double exponential model~\cite{mcguire_palmer}, including both a rise time and a decay time. Two further decaying exponentials are included for LAB/PPO, based on~\cite{labppo}. The simulated time profile is:
\begin{eqnarray}
\rho(t) \propto (1 - e^{-t/\tau_r}) \times \sum^3_i A_i e^{-t/\tau_i},
\end{eqnarray}
where $\tau_r$ is the rise time and the parameters $\tau_i$ and $A_i$ are the decay times and their scale factors.

\subsection{Optical Propagation \label{sec:optics}}

Optical photons are propagated through different media by GEANT4. Absorption, refraction and reflection are taken into account according to the material's optical properties. The absorption length of the UVT acrylic used in CHESS (\Cref{fig:optics}) has been measured with a spectrometer and is included directly in the simulation.  
The refractive indexes for LAB and LAB/PPO are considered to be the same and are taken from~\cite{snop_private}. \\
Photon reemission from wavelength shifting materials are simulated in a similar fashion as the scintillation process. Once a photon is absorbed in a medium, there is a finite probability that it will be reemitted following a specified reemission spectrum.

\subsection{PMT Model \label{sec:pmt_model}}

A full and precise simulation of the different PMTs is implemented in RAT-PAC. It contains a detailed PMT geometry (glass, photocathode, dynode stack and case) as well as a dedicated photon tracking simulation inside the PMT volume. Once an optical photon hits the external boundary of the PMT glass, it is propagated through the different internal PMT surfaces according to the relevant material optical properties.  The QE of the PMTs is taken from Hamamatsu specifications~\cite{h11934} and used as an input to the simulation to determine whether or not to create a PE for an incident photon at a particular wavelength. An individual normalization can be applied to the efficiency of each tube to allow for  
a finite collection efficiency.  These are set to 90\% for each tube~\cite{hamamatsu}.  
If the incident photon creates a PE, its associated charge and time are extracted from Gaussian distributions that have been previously calibrated to take into account the PMT gain and electronic delays (\Cref{sec:calibration}).

\subsubsection{Charge Distribution Model}

The single-PE (SPE) charge distribution is modeled as a Gaussian truncated for negative charge values. The SPE charge distribution for individual PMTs has been measured (\Cref{sec:calibration}) and \Cref{fig:spe_data} shows that the Gaussian model agrees well with the data.

\subsubsection{Time Profile Model}

Each PMT has a characteristic transit time and TTS that depends on the PMT design. These numbers are taken from the PMT specifications and included in the simulation assuming the time is Gaussian distributed. The PMT specifications~\cite{h11934} show that the Gaussian model is a very good approximation to the time profile. The mean is set to the provided transit time and the width to the provided TTS. Extra time offset parameters allow for individual delays per electronics channel (referring to the PMT, cable, and readout electronics) primarily due to different cable lengths. These are measured on a channel-by-channel basis (\Cref{sec:calibration}).

\subsection{DAQ Simulation \label{sec:daq_sim}}

For each event an analog waveform is generated per channel by summing the individual pulses created by each PE (\Cref{sec:pmtpulses}). 
The waveform is then digitized via a process that mimics the characteristics of the two models of CAEN digitizers used in the detector. 
High frequency electronics noise is added to these digitized waveforms following a Gaussian distribution centered at zero and with a width of $0.35$~mV and $0.88$~mV for the V1730 and V1742 cards, respectively.
These widths are measured from data by performing a Gaussian fit to the residuals of pedestal-corrected noise data.
In this way, any effects introduced into the dataset by the digitization process are reproduced in the simulation.

The trigger process implements the conditions described in \Cref{sec:triggering} and decides whether to create a triggered detector event. 
When an event is created, acquisition windows corresponding to the buffer size of the appropriate digitizer are captured from the digitized waveforms.
These windows are used to determine photon hit times and charge using the same process as applied to data in Sec.~\ref{s:analysis}.  
The trigger process then scans the remainder of the digitized waveform to look for additional triggers within the same simulated physics event.

\section{Waveform Analysis}\label{s:analysis}

The full waveform is analyzed to extract individual PMT charges and hit times. 
An analysis window of $135$~ns (675 samples) is chosen starting $160$~ns (800 samples) prior to the acquisition trigger. This is defined based on when light from the target is expected to hit the PMT array. The charge in picocoulombs is defined as the integral within that window multiplied by a 50~$\Omega$ resistance, corrected by the pedestal charge. The pedestal charge is calculated on a trace-by-trace basis by taking an average of sample values across the pedestal region, which is defined by a $25$~ns window right before the analysis window. If fluctuations of more than $5$~mV peak-to-valley are detected in the pedestal region, the waveform is not analyzed.  The total integrated charge is converted to an estimated number of detected PEs by normalizing by the mean value of the SPE charge distribution measured in \Cref{sec:calibration}.

The time of an individual PMT pulse is defined as the time at which the waveform crosses a threshold given by $25\%$ of the  height of a single PE pulse. 
This threshold is determined by modeling the PMT pulse as a log-normal pulse with parameters extracted from the fit in \Cref{sec:pmtpulses} and an integrated charge equivalent to the mean of the SPE distribution for each PMT (\Cref{pmt-gain}). 
This threshold is well clear of the per-channel noise levels, as can be seen in Fig.~\ref{fig:spe-pulse-shape}.

Linear interpolation between ADC samples is used to maximize the time precision. The measured hit time at the PMT array is corrected by the photon time of flight (ToF), assuming that the photons are produced in the center of the target. 

\section{Calibration}
\label{sec:calibration}

The setup is calibrated using a $^{90}$Sr beta source in combination with a water-filled target, an LED deployed in the dark box, and a control sample of cosmic muons passing through the acrylic propagation medium. For source data, the veto panels are used to reject coincident cosmic muon events. Beta decay, muon ionization and Cherenkov emission are well understood processes that help to calibrate the optical aspects of the detector. 
The following sections detail the calibration techniques.

\subsection{PMT Gain}
\label{pmt-gain}
\begin{figure}
	\centering
	\includegraphics[width=\columnwidth]{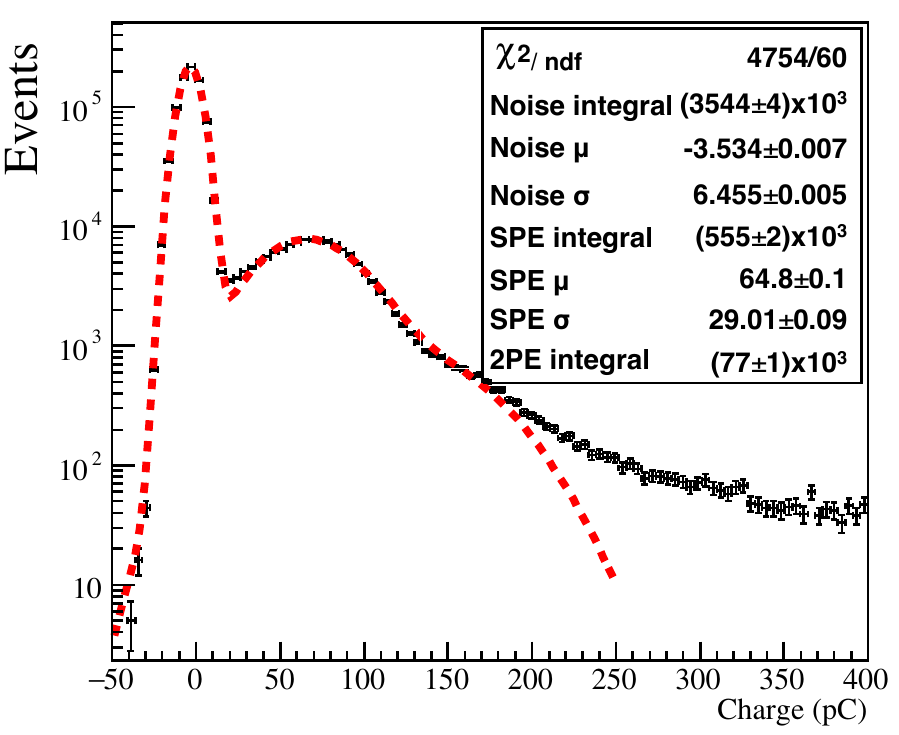}
	\caption{SPE charge distribution for a single PMT. Data is shown as black points with error bars, and the red dashed line is the result of the fit used to extract the shape of the SPE distribution. }
	\label{fig:spe_data}
\end{figure}

PMT gains depend on the supplied voltage and vary tube-by-tube, so the gain is measured individually for each PMT in the array. The PMT voltages are kept constant between data-taking periods. A direct measurement of the gain is given by the SPE charge distributions. A dim source of light is provided by the $^{90}$Sr beta source (Sec.~\ref{s:source}) attached to the water target. 
A small PMT optically coupled to the target is used to trigger the acquisition (Sec.~\ref{sec:triggering}). The SPE charge distribution for one of the array PMTs  is shown in \Cref{fig:spe_data}, where a clear SPE peak is identifiable. There is a noticeable population of multi PE events, 
which may be due to byproducts of muon cosmic events crossing the propagation medium. The muons themselves are easily vetoed using the scintillator panels (Sec.~\ref{s:veto}), but high energy electrons and gammas produced by these muons are more difficult to veto and thus cause irreducible tails in the charge distribution. 

A Gaussian fit is used to model the charge distributions. To precisely extract the SPE parameters, the noise distribution and the multi-PE event distributions with up to 3 PEs are included and re-weighted by nuisance parameters. A Gaussian is assumed for the noise peak distribution with a mean, width and integral floated in the fit. For multi-PE events only the event rates with respect to SPE are floated in the fit, since the mean and width are determined by the SPE distribution. As a result, a fit with 7 parameters (SPE mean and width, noise mean, width and integral, multi-PE event rates for 2 and 3 PEs) driving a multi-Gaussian model is performed and the mean and width for the SPE of each single PMT is extracted.  \Cref{fig:spe_data} shows a sample fit. 
Before inclusion in the Monte Carlo model the SPE width is corrected for noise, since this is independently modeled, 
by subtracting in quadrature the width of the noise peak from the fitted SPE width.

The stability of the PMT gains was checked by taking a second set of water calibration data after LS data taking was complete.  The gains for all PMTs in the array were observed to be very stable within measured uncertainties.

\subsection{PMT Pulse Shapes}
\label{sec:pmtpulses}
\begin{figure}[!t]
	\centering
	\includegraphics[width=\columnwidth]{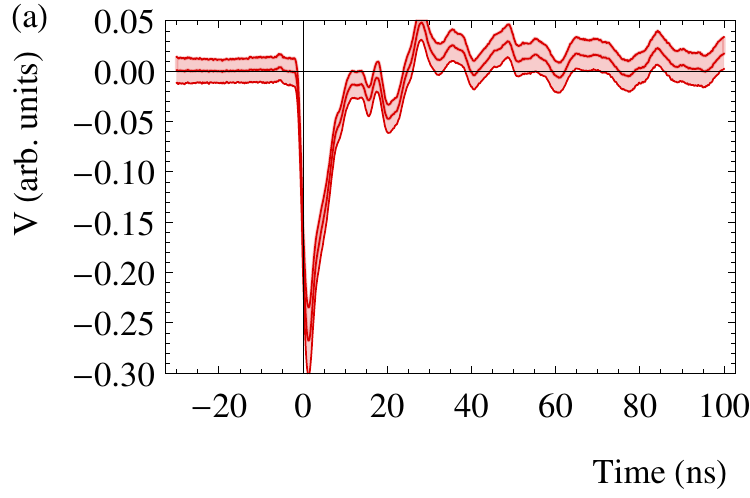}
	\includegraphics[width=\columnwidth]{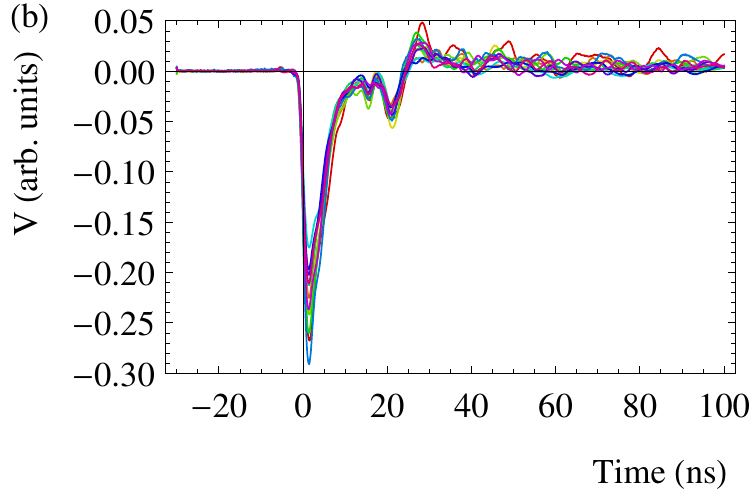}
	\includegraphics[width=\columnwidth]{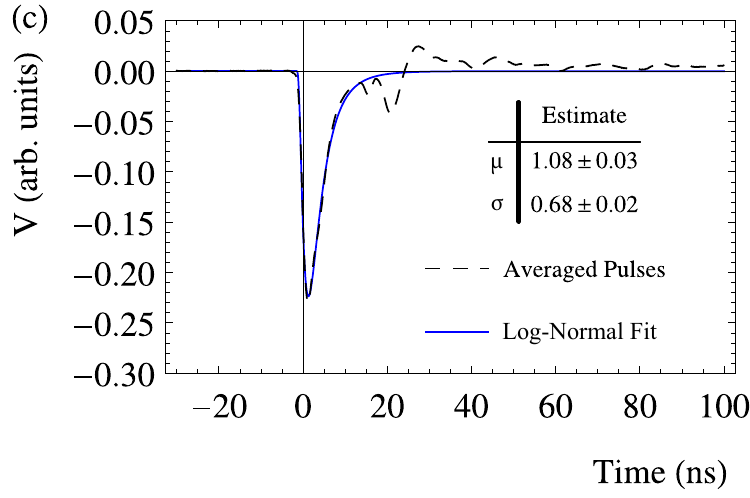}
	\caption{PMT pulse shape characterization.   (a) 
	Mean and RMS spread for one PMT, showing excellent stability in the pulse region.  (b) Averaged pulse shapes for SPE pulses for each individual PMT. 
	(c) Log-normal fit to the average pulse shape across all PMTs. 
	In all cases the pulses are normalized to unit area, so the voltage is reported in arbitrary units.}
	\label{fig:spe-pulse-shape}
\end{figure}

The characteristic SPE pulse shapes need to be well modeled in order to accurately reproduce the PMT time response in simulation. 
Events with an integrated charge from $-\frac{\sigma}{2}$ to $+\sigma$ about the mean of the SPE charge were extracted from the SPE calibration data  for each PMT independently. The smaller bound on the negative range was introduced to avoid including noise events. These pulses were then normalized to unit area and a constant fraction threshold was applied to determine a common point of reference between all pulses from a single PMT.  
This threshold was used to align pulses on a per-PMT basis, and a mean value along with upper and lower RMS values were calculated for the normalized voltage in 100~ps time bins. The result is shown for one PMT in the top panel of \Cref{fig:spe-pulse-shape}. The uncertainty in the pulse region is consistent with the noise of the ADCs; this was true for all PMTs. The extracted mean PMT pulse shapes are shown for all 12 array PMTs in the middle panel of \Cref{fig:spe-pulse-shape}. The PMT-to-PMT variation in pulse shape is attributed to small variations in voltage dividers, signal coupling circuits, and individual PMT construction. A log normal of the form 
\begin{eqnarray}
\frac{1}{\sigma \sqrt{2\pi}}\exp{\left[\frac{-\left(\log[t-t0]-\mu\right)^2}{2\sigma^2}\right]}
\end{eqnarray}
was fit to the average pulse shape across all 12 PMTs (lower panel of \Cref{fig:spe-pulse-shape})
and the resulting fit parameters were used in the simulation to model PMT pulse shapes.
	
\subsection{Per-Channel Time Delays}\label{s:delay}

The total time delay due to PMT transit times, electronics, and cable lengths is measured for each individual channel and then corrected for in the data. 
Time delays are measured relative to the average across all PMTs in the array. 
Typical relative delays are on the order of 100s~ps. 
The primary calibration is performed by deploying an LED at the top of the dark box, such that the light path to each PMT is similar.  A full Monte Carlo simulation of this configuration is run and used to correct the data in order to take into account PMT-to-PMT variations in photon ToF and geometry effects.  
Several datasets are generated by removing and reattaching the LED in order to quantify any uncertainty in the LED position with respect to the simulated one. 
Cosmic muons passing through the propagation medium are used to evaluate systematics uncertainties in the calculation of the time delays, as they provide an abundant source of prompt Cherenkov light. Events are selected  by requiring a coincidence between the lower muon tag and the bottom scintillator panel. Again, a full Monte Carlo simulation is used to correct for ToF and geometry effects in the calibration.  

\begin{figure}
	\centering
	\includegraphics[width=\columnwidth]{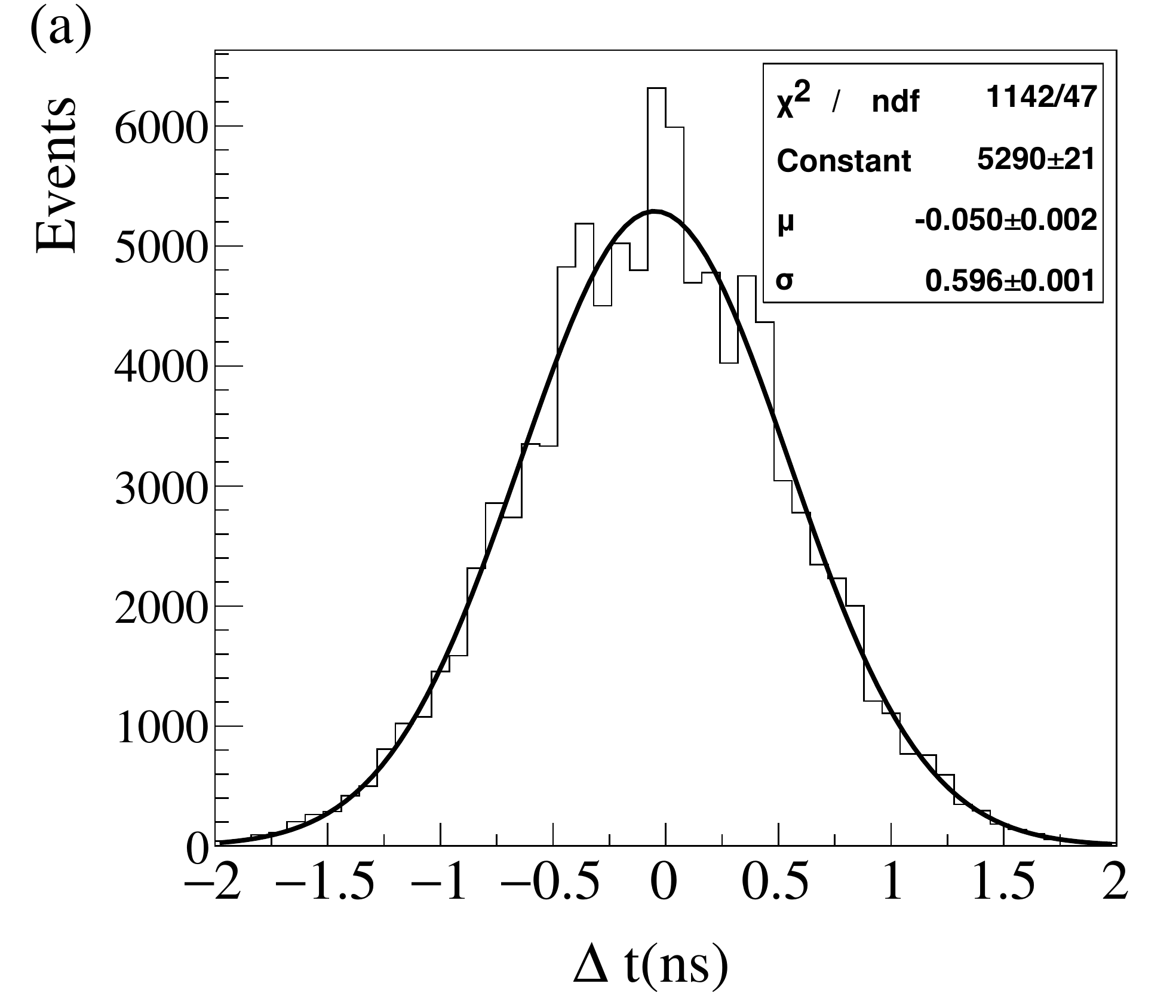}
	\includegraphics[width=\columnwidth]{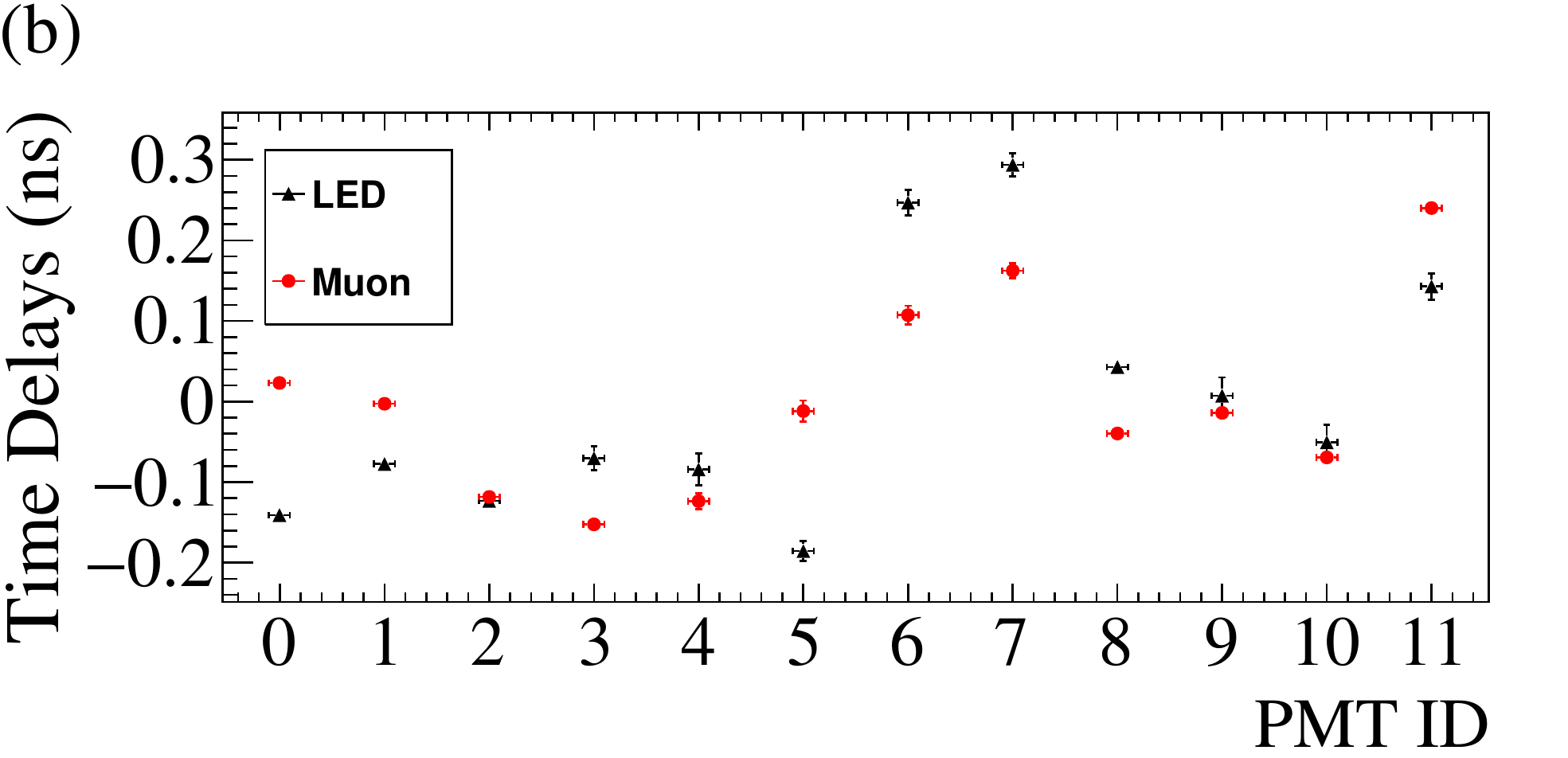}
	\caption{(a) Time distribution for a single channel relative to a reference channel, overlaid with the Gaussian fit used to extract the relative time delay.  (b) Mean and uncertainty on the mean for the per-channel time delays.}
	\label{fig:time_delay}
\end{figure}

The time distribution for one channel is shown in the upper panel of \Cref{fig:time_delay}, where the offset represents the time delay with respect to the average. 
The measured time delays with uncertainties for each channel from each calibration are shown in the lower panel of \Cref{fig:time_delay}.  The two data sets show the same trend across the PMT array.  
The MC-corrected LED data set is used to define the time delays for the final separation analysis.  The difference between the two data sets is taken as a measure of  systematic error in this calibration.

\section{Cosmic Muon Event Selection}\label{s:event}

A deionized water target was used to optimize  event selection criteria in order to select Cherenkov ring events, and to reject backgrounds caused by instrumentals and by other physics events.  These criteria are then used to determine the sensitivity to identifying ring candidates in LS.  
The criteria were defined by classifying all events in the water data set according to the clarity of a visible Cherenkov ring (Sec.~\ref{s:class}), and then adjusting relevant cut values to maximize acceptance of good ring candidates and minimize contamination by non-Cherenkov-like events (Sec.~\ref{s:cut}).

\subsection{Event Classification}\label{s:class}
Events in the pure water data set were hand scanned in order to understand the different types of event topology and their sources.  A classification scheme was developed to sort them according to how well they matched 
the expected Cherenkov ring geometry, for the purposes of defining a quantitative set of event-selection criteria (Sec.~\ref{s:cut}). PMTs with an integrated charge greater than one third of the SPE charge for that PMT were counted as hit.  Hits were grouped according to the radius of the hit PMT.  The cross-shape PMT deployment results in three radial groupings: the {\it inner}, {\it middle} and {\it outer} PMTs.  The CHESS apparatus was designed such that the Cherenkov ring falls on the middle ring for a water target, and the outer ring for a pure LS target.  The total number of hit PMTs within each grouping, 
NHit$_{inner}$, NHit$_{middle}$, and NHit$_{outer}$, was determined by summing hits across all PMTs within that group. A perfect Cherenkov ring in water is expected to have NHit$_{middle} = 4$ while NHit$_{inner} = {\rm NHit}_{outer} = 0$. ``Ring'' events were selected with the criteria NHit$_{middle} - {\rm NHit}_{inner} - {\rm NHit}_{outer} > 2$ to allow either one expected PMT to be missed or one additional PMT to be hit (but not both) to allow for minor noise contamination and increased acceptance.

``Background'' events were sorted into categories according to  event topology, in order to understand the primary sources of background.  These included instrumental events, which had no clear ring, so-called ``follower'' events, in which a secondary particle generated Cherenkov light in the propagation medium, causing unusually high charge on the inner PMTs, and events in which a cosmic muon shower lit up the majority of PMTs within the array.

\subsection{Cut Development}\label{s:cut}
Selection of vertical-going cosmic muon events was achieved using a triple coincidence trigger.  The hardware trigger threshold (Sec.~\ref{sec:triggering}) applied to the lower muon tag was set conservatively low to maximize event acceptance.  A software trigger was applied offline by applying a threshold to the charge on both the upper and lower muon tags and to the muon panel directly below the setup in order to select coincidence events.  

Rejection of events containing either multiple muons or secondary particles was achieved by requiring the charge on all veto panels except the panel directly below the setup to be consistent with zero.  

\begin{figure*}
\centering
\begin{tabular}{cc}
\includegraphics[scale=0.4]{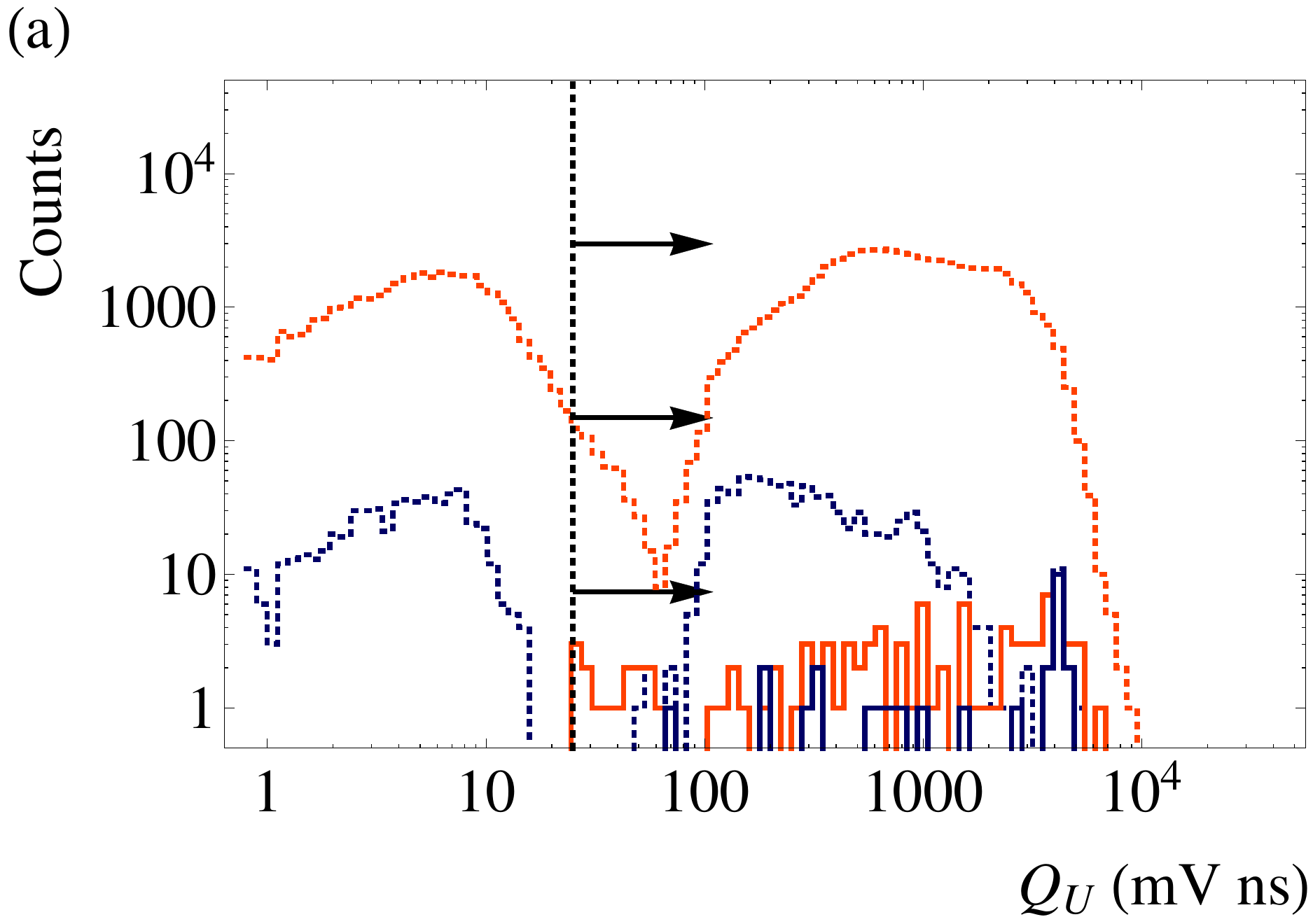} &
\includegraphics[scale=0.4]{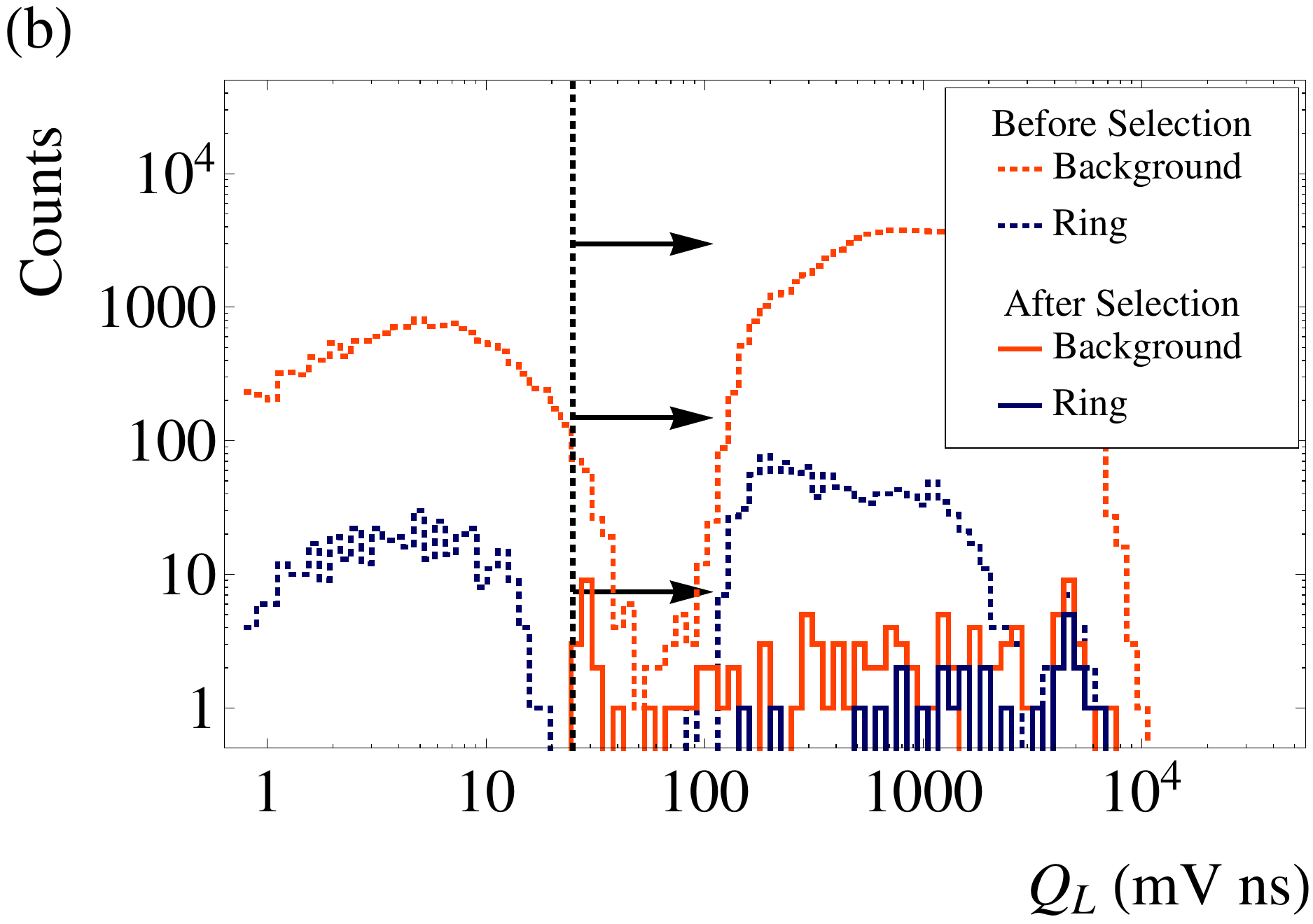} \\
\includegraphics[scale=0.4]{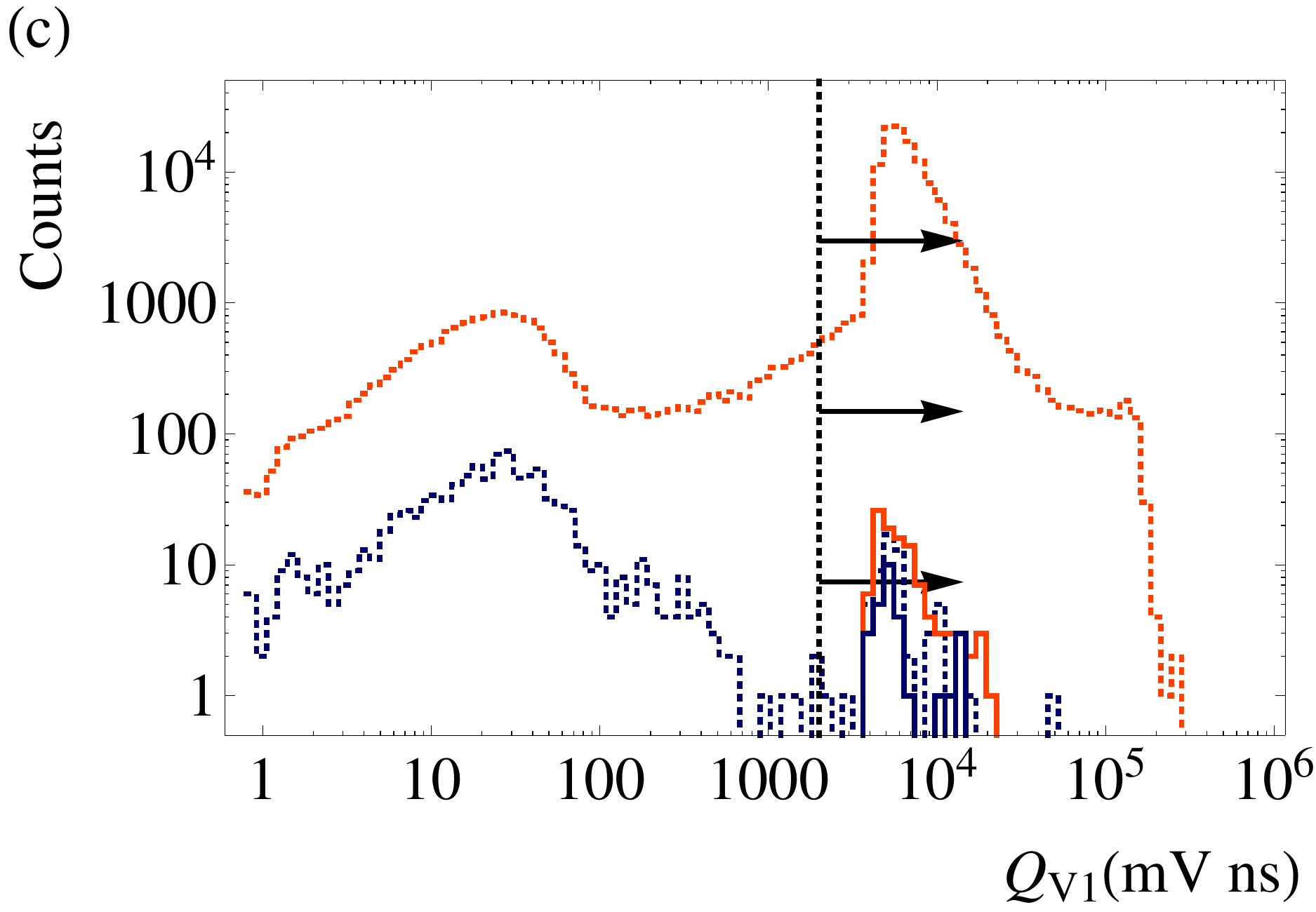} &
\includegraphics[scale=0.4]{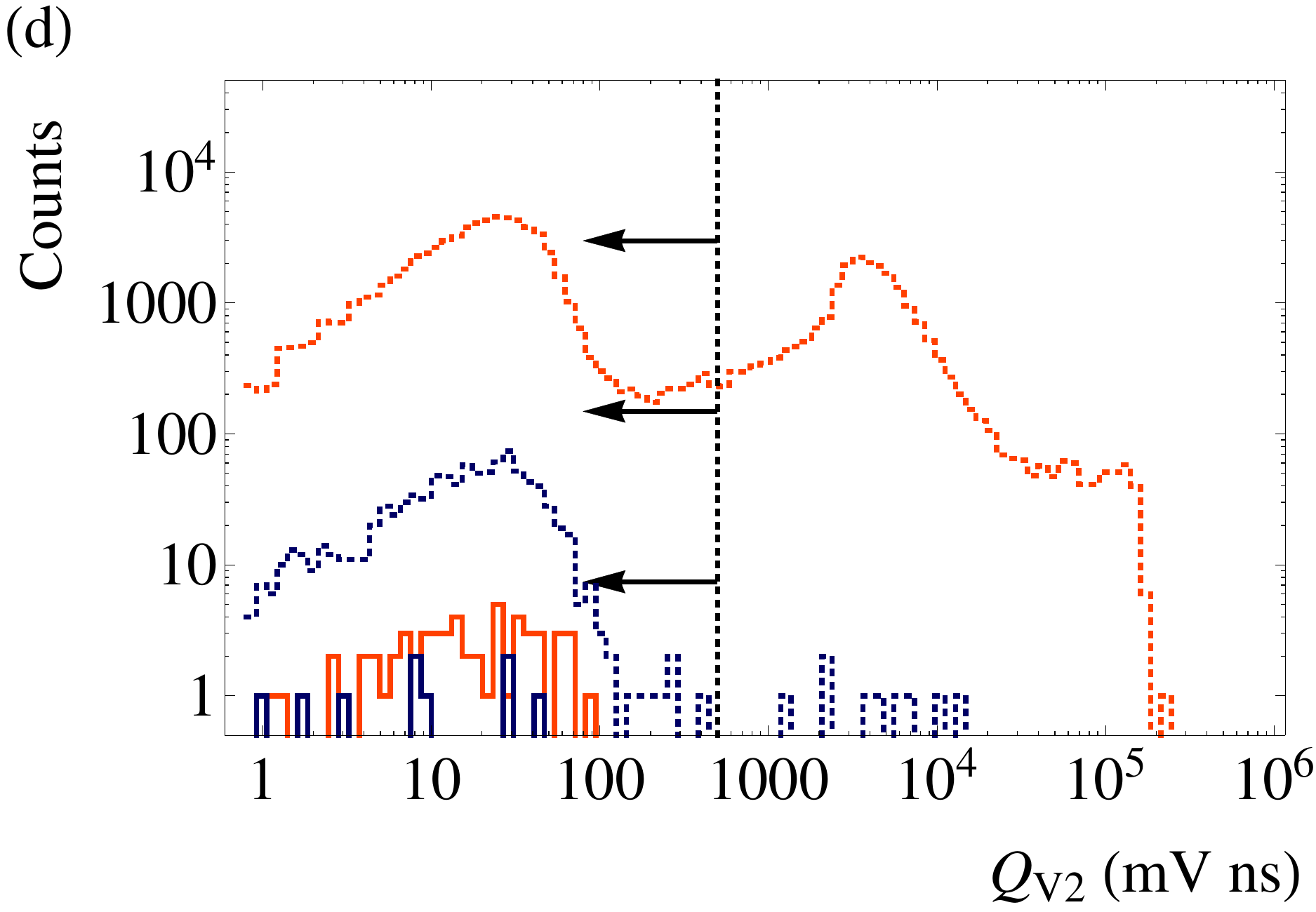} \\
\includegraphics[scale=0.4]{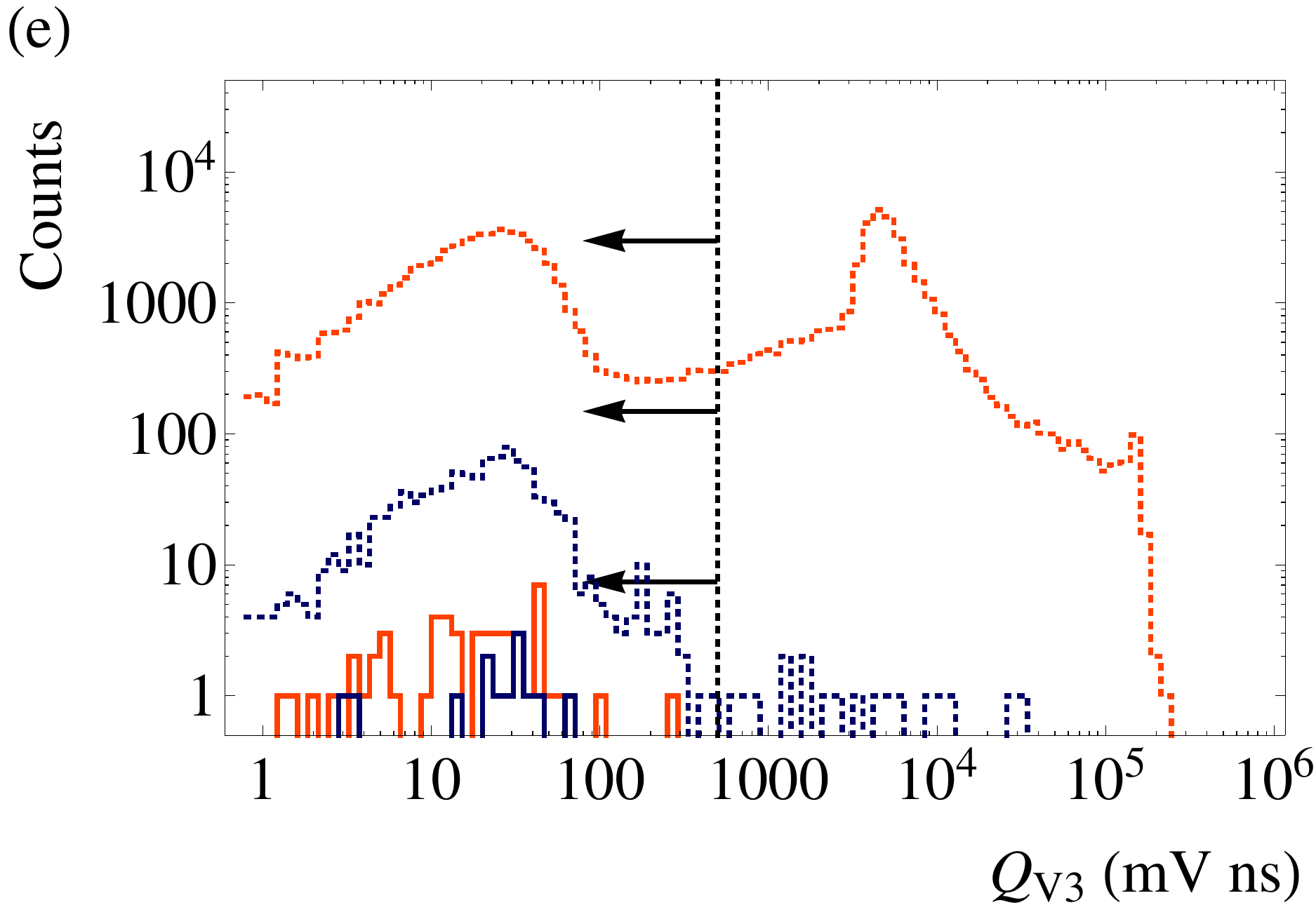} &
\includegraphics[scale=0.4]{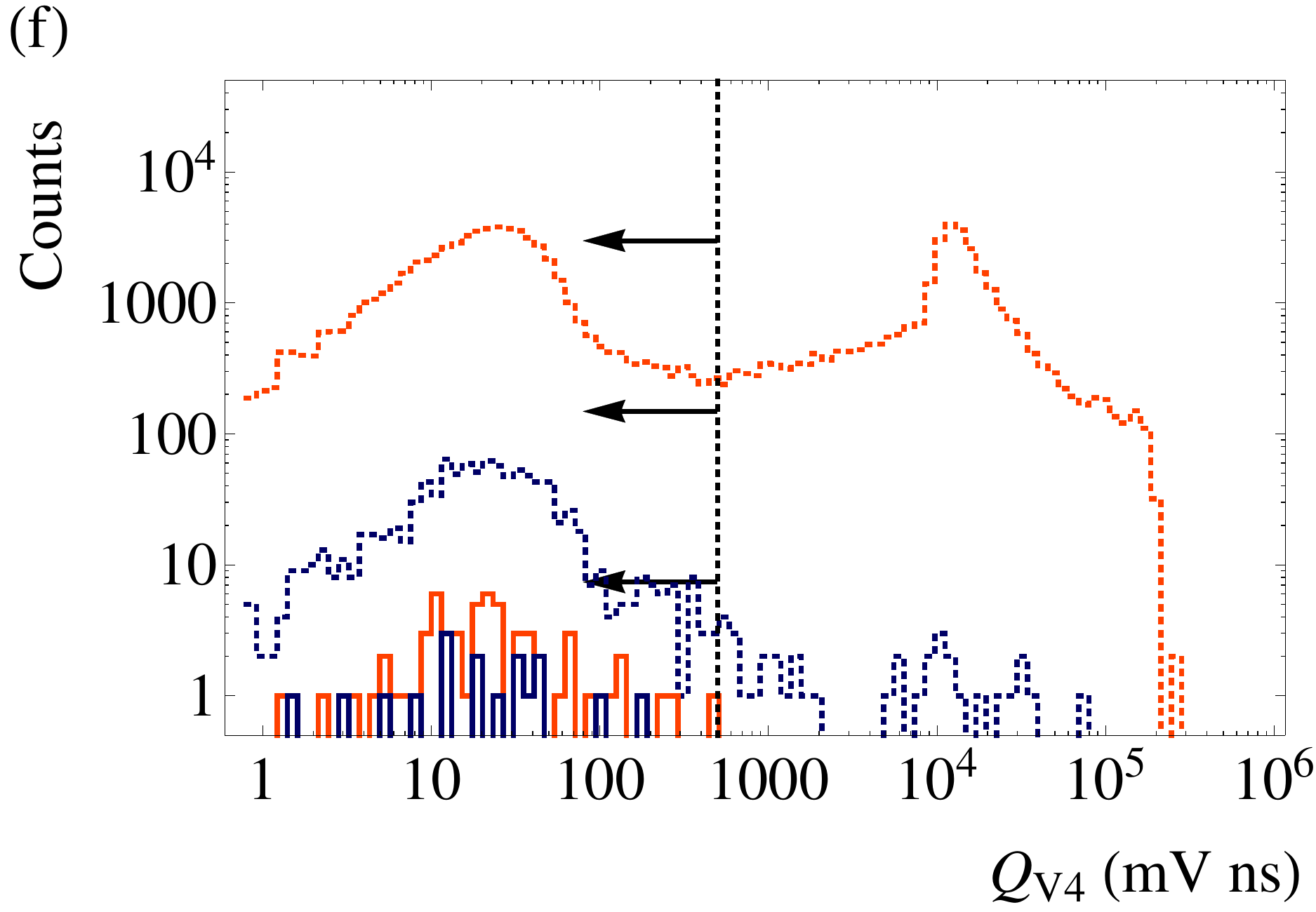} \\
\end{tabular}
\caption{(Top to bottom, left to right) Charge distribution of events on the upper and lower cosmic tags ($Q_U$ and $Q_L$) and the four muon panels ($Q_{V1}$ -- $Q_{V4}$).  Panel 1 is located directly below the CHESS apparatus.  Events are separated into ring (blue, lower line) and background (red, upper line) according to the criterion in Sec.~\ref{s:class}.  Vertical black dashed lines show the cut values in each case with arrows indicating the acceptance region.  Distributions are shown before (dashed) and after (solid) application of cuts on these 6 parameters.}
\label{fig:event-selection-cuts}
\end{figure*}

The threshold values applied to the charge seen on the cosmic tags ($Q_U$ and $Q_L$ for the upper and lower tag, respectively), and for each of the muon panels ($Q_{V1}$ -- $Q_{V4}$) were selected by optimizing acceptance of ``ring'' events and rejection of ``background'' events, according to the classification described in Sec.~\ref{s:class}.  The data before and after application of these cuts is shown in \Cref{fig:event-selection-cuts}.

\begin{figure}
\centering
\includegraphics[width=\columnwidth]{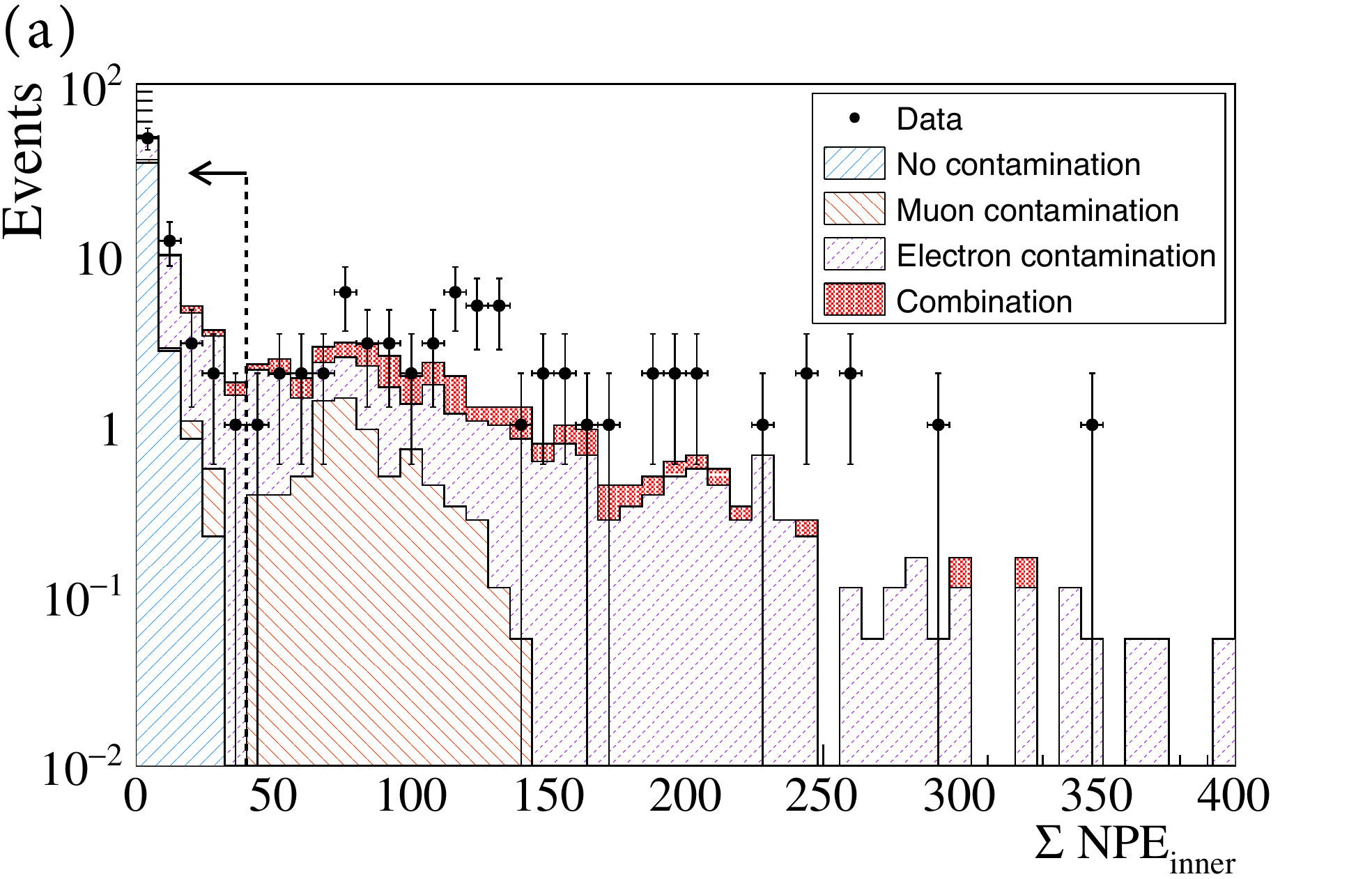}
\includegraphics[width=\columnwidth]{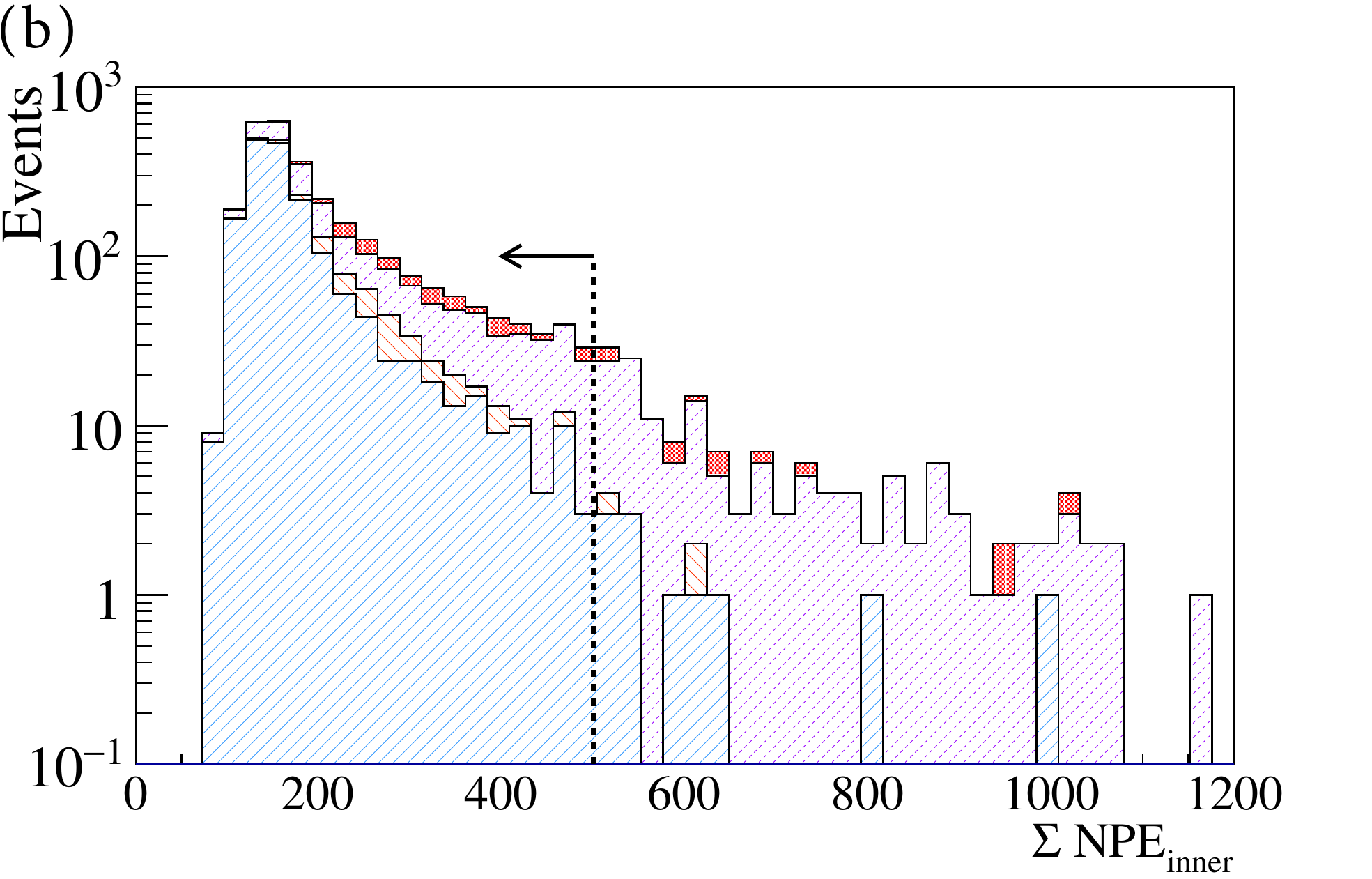}
\caption{ Monte Carlo simulation of the summed PE distribution on the inner PMT group due to cosmic muon events with perfect rings (turquoise, right diagonal hatching), muon contamination (orange, left diagonal hatching), electron contamination (purple, right diagonal dashed hatching), and the total (red, solid).  (a) Water target, with water data overlaid to show the agreement, and (b) LAB/PPO target.  The vertical black dashed line represents the chosen cut value, with arrows to illustrate the acceptance region.}
\label{f:clipMC}
\end{figure}

Cuts were also applied to remove so-called ``follower'' events, in which muons or muon followers generated Cherenkov light in the acrylic propagation medium, which contaminated the sample of single Cherenkov ring events.  This occurred in two cases: 
\begin{itemize}
\item \emph{Electron contamination:} The cosmic muon triggered the acquisition, but a secondary particle passed through the propagation medium.
\item \emph{Muon contamination:} The secondary particle triggered the lower muon tag and the muon itself passed through the propagation medium.  
\end{itemize}
The secondary particles do not always make it to the muon panels and therefore cannot be vetoed directly, thus the only information available for rejecting these events is the PMT array.  
Clean cosmic muon events are expected to produce hits only on the middle PMTs (for a water target, or outer PMTs for LS).
In both cases of follower events, the majority of Cherenkov light generated in the propagation medium was observed 
to fall primarily on the innermost PMTs within the array.  This was confirmed by Monte Carlo simulations of each event type, which 
demonstrated that muon followers do produce events with this topology.   
Simulations of these events show a clear tail in the PE distribution observed on the inner PMT group for both water and LAB/PPO targets (Fig.~\ref{f:clipMC}).  
These events 
typically create between 30 and 400 PEs in the innermost PMTs, making their identification in both water and LS possible by analysis of the charge on these PMTs.  

A cut was designed to reject these events based on event topology: since the Cherenkov ring geometry is not expected to produce hits on the inner PMTs for either a water or LS target, the total number of estimated PEs on the inner PMTs was used to identify clean rings.   In water this charge is expected to be very low (consistent with noise), whereas in LS the total charge will be higher due to scintillation photons.  However, as demonstrated in Fig.~\ref{f:clipMC}, a large fraction of follower events can still be removed with a high-charge cut.   
A cut was placed at a summed PE count of 40 for water and LAB targets, and 500 for LAB/PPO.  
These cuts were conservatively chosen based on the simulations to reject events with high light contamination due to secondary particles with a minimal impact on the efficiency. 
Performance of this cut in water is shown in 
the top panel of Fig.~\ref{f:clipMC}.  The high-charge tail due to follower events is similar to that seen in the simulation, supporting use of the simulation for defining this cut for both water and LS targets. The  cut value selected for a water target removes a large fraction of the remaining background events, with zero sacrifice in the control sample.  

\section{Event-Level Analysis}\label{s:recon}

Time separation of the Cherenkov and scintillation photon populations is based on the hit-time residual distributions for each radial PMT grouping (inner, middle and outer).  The hit-time residuals are evaluated as the PMT hit times with respect to the event time, corrected by per-channel delays (Sec.~\ref{s:delay}) and by the photon ToF.  
The ToF depends on the distance between the target and the PMTs, and on the refractive indices of the different media. It is estimated for each PMT radial group to be $626$~ps, $536$~ps and $473$~ps for the inner, middle and outer PMT rings, respectively.

The time at which the cosmic muon passes through the target, referred as the event time, is calculated using two different approaches. The most straightforward is using the time at which the lower muon tag goes above threshold, since this  triggers the acquisition. Nevertheless, the cosmic tags suffer from a poorer time resolution than the fast PMTs in the array due both to the scintillator response and the larger PMT TTS. 
Hence, a higher precision event time is defined using the PMT hit times by using the median of the four earliest hits in the event, after time calibration and photon ToF correction. This provides a robust time reference since the prompt hits are due to Cherenkov light, whose time profile is very sharp.  

\begin{figure}
	\centering
	\includegraphics[width=0.9\columnwidth]{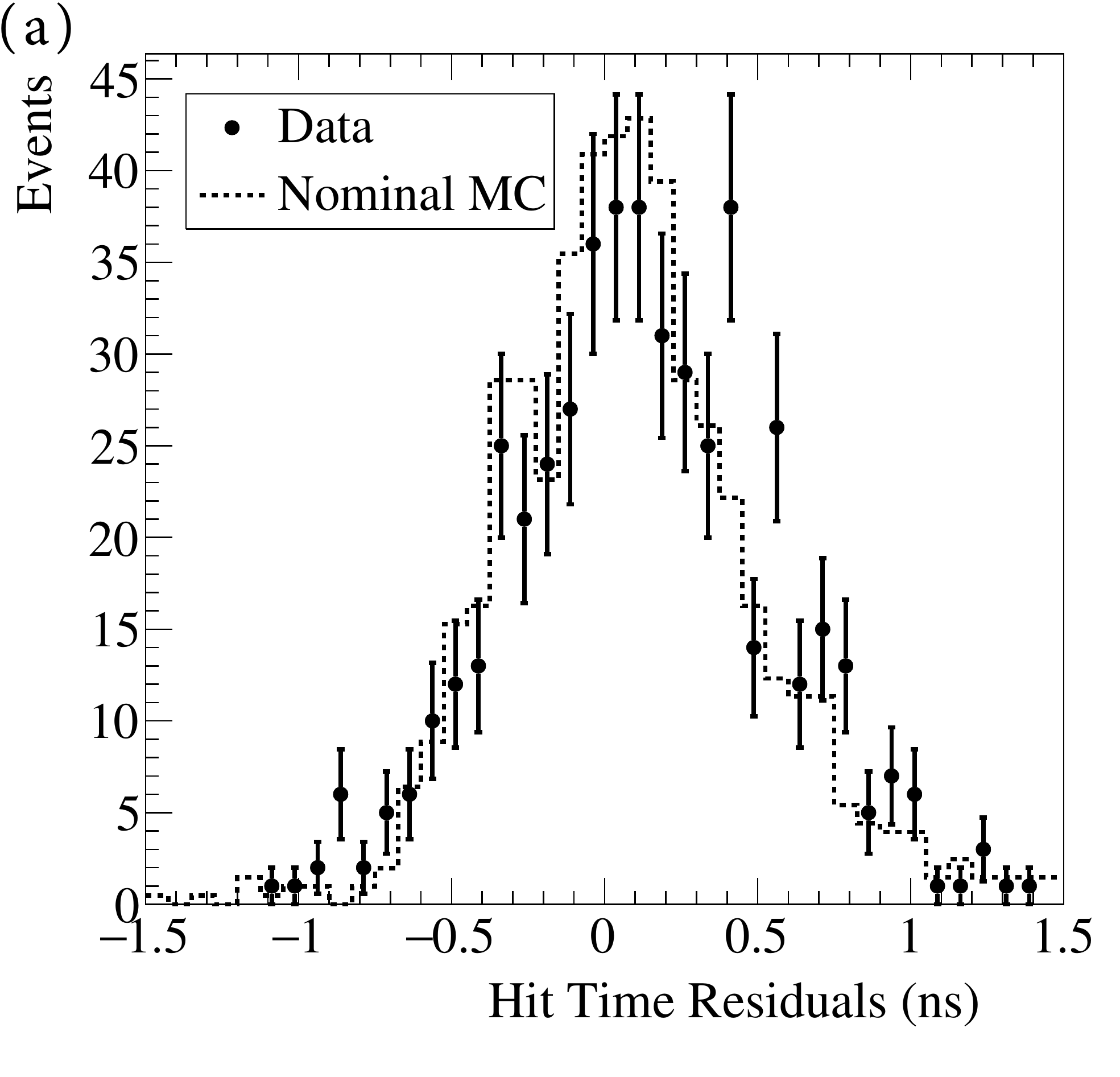}
	\includegraphics[width=0.9\columnwidth]{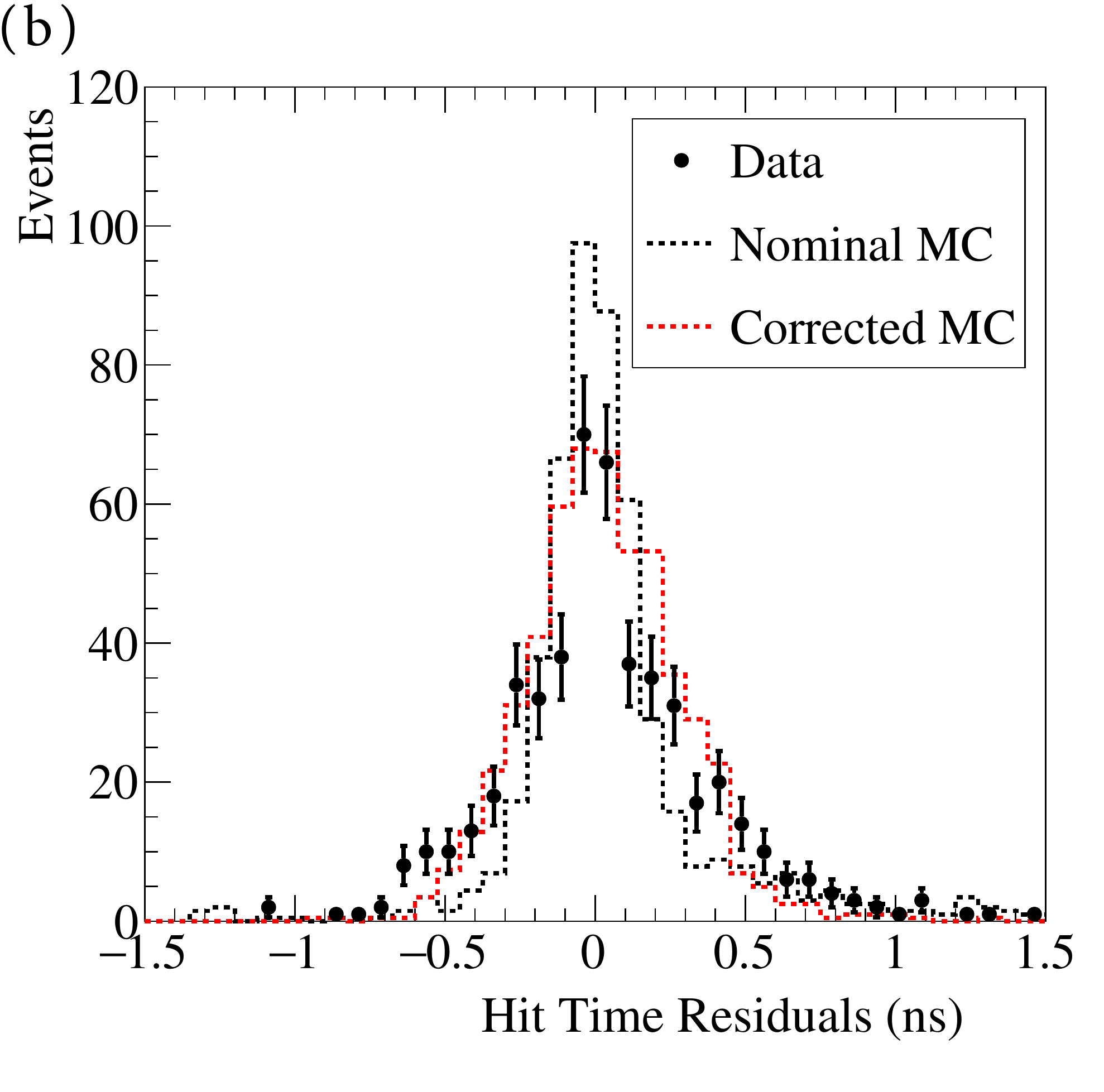}
	\caption{Distribution of hit-time residuals 
	for ring candidate events in water for middle PMTs, where the Cherenkov ring is expected. Data points are shown with statistical errors, with the Monte Carlo prediction overlaid (dashed lines). 
	(a) Hit times with respect to the lower muon tag time and (b) hit times with respect to the reconstructed event time.}
	\label{f:smear}
\end{figure}

The hit-time residual distributions evaluated using each option for the event time are shown in Fig.~\ref{f:smear}, 
overlaid with the Monte Carlo prediction in each case.  
The residual distribution with respect to the lower muon tag time is extremely well reproduced by the Monte Carlo, demonstrating the precision of the model.  The simulation slightly under predicts the width of the higher precision distribution of residuals with respect to the reconstructed event time.  This could be explained by multi-photon effects, small differences in the PMT pulse shape, and underestimated PMT TTS.  
A Gaussian correction of $\sigma=214$~ps is included in the Monte Carlo in order to take these effects into account, and to better reproduce the time resolution seen in data. 

\section{Cherenkov Ring Imaging in Water}

\begin{figure}
	\centering
	\includegraphics[width=7.95cm]{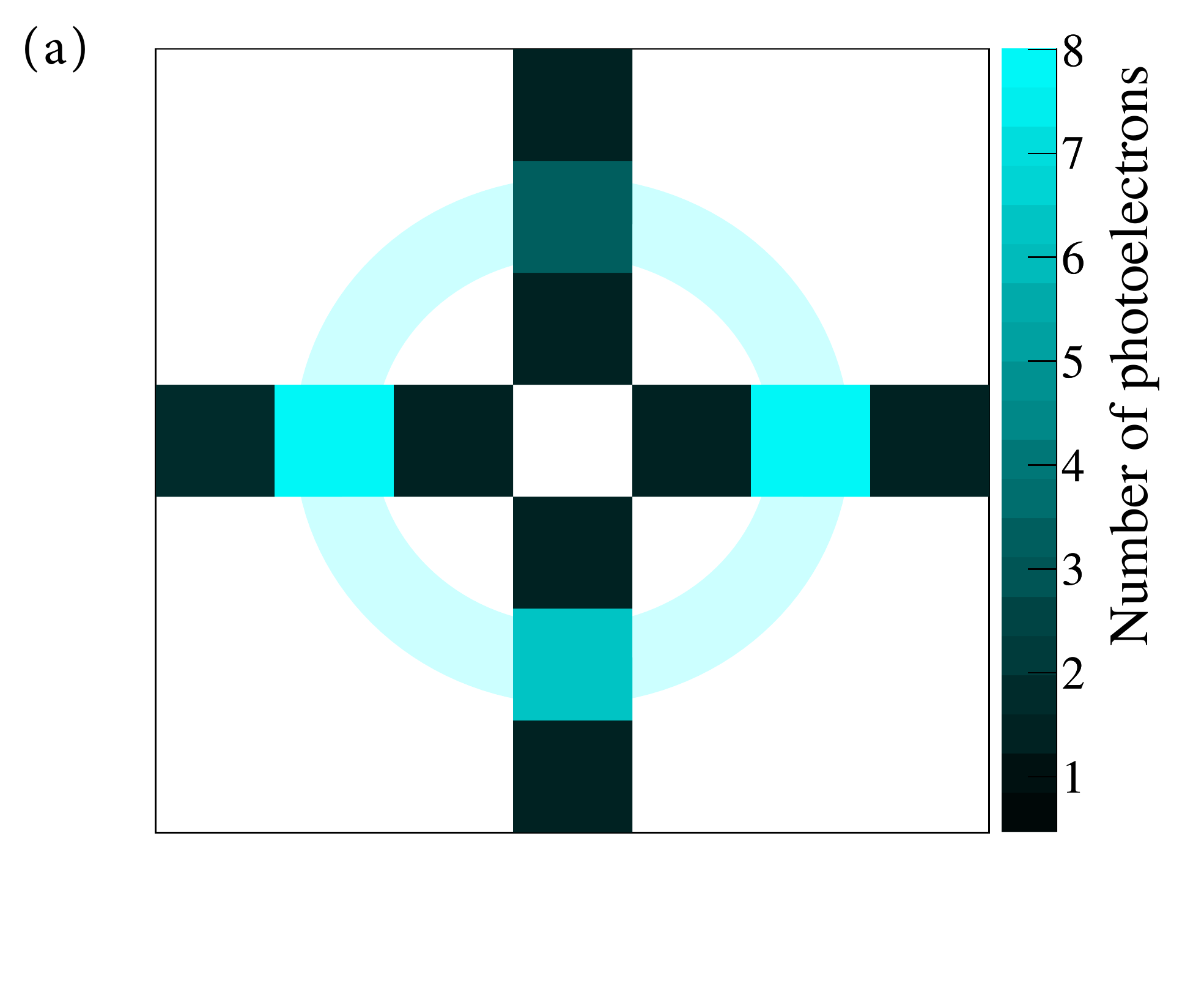}
	\includegraphics[width=8cm]{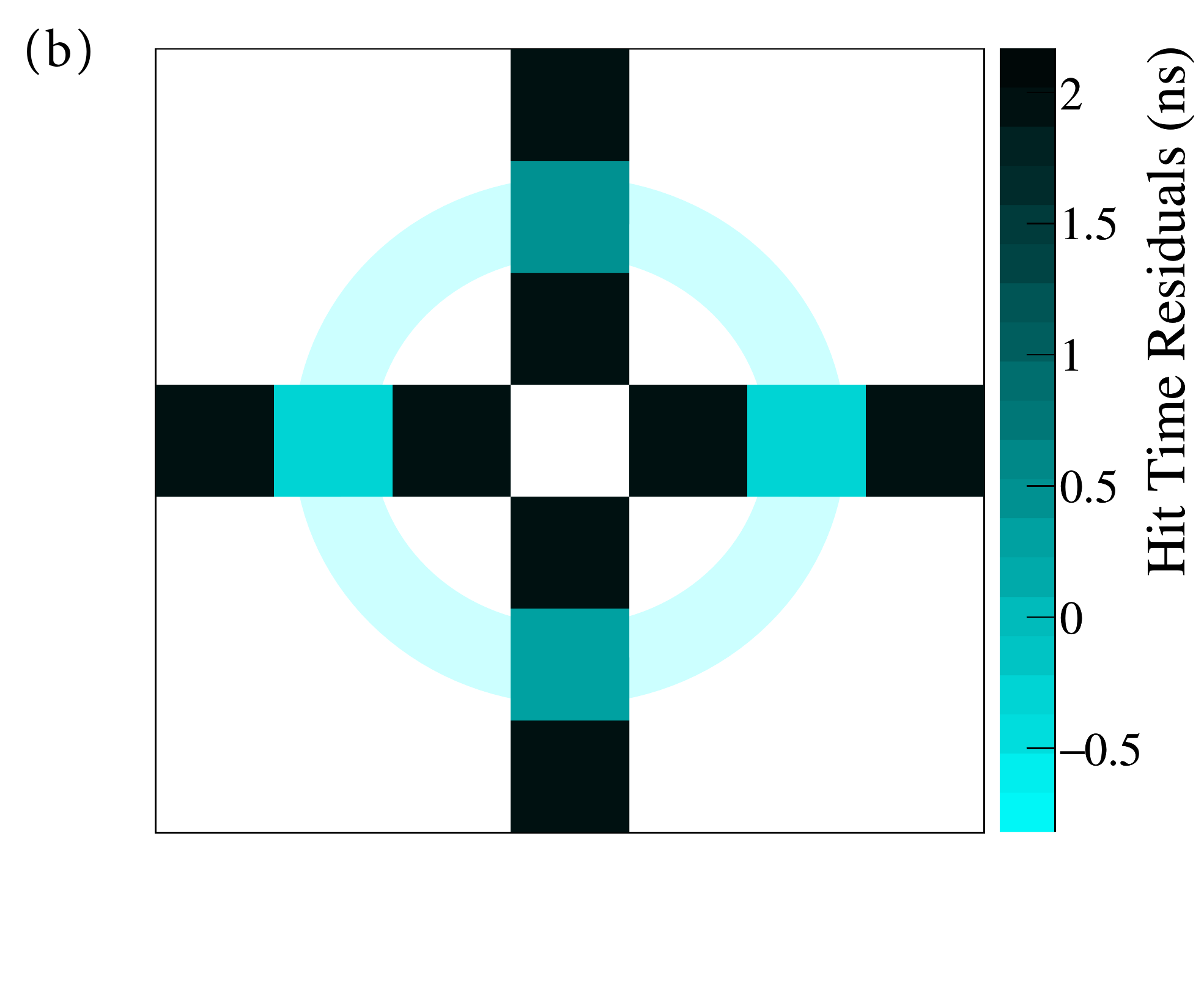}
	\caption{Typical ring event in the water data set. (a) Estimated number of detected PEs and (b) hit time residuals.}
	\label{fig:cosmics_water_ring_candidate}
\end{figure}

\begin{figure}
	\centering
	\includegraphics[width=7.8cm]{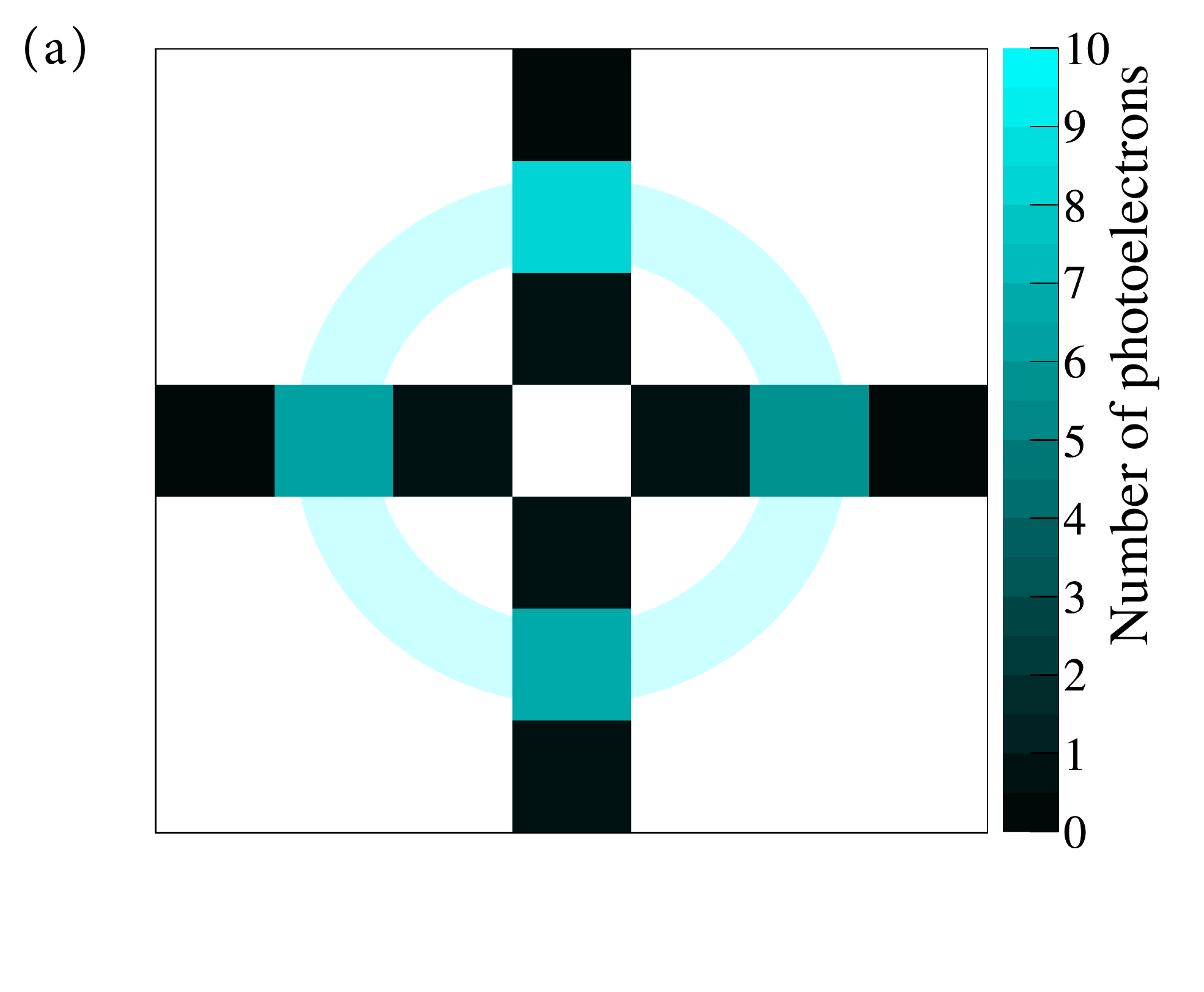}
	\includegraphics[width=8cm]{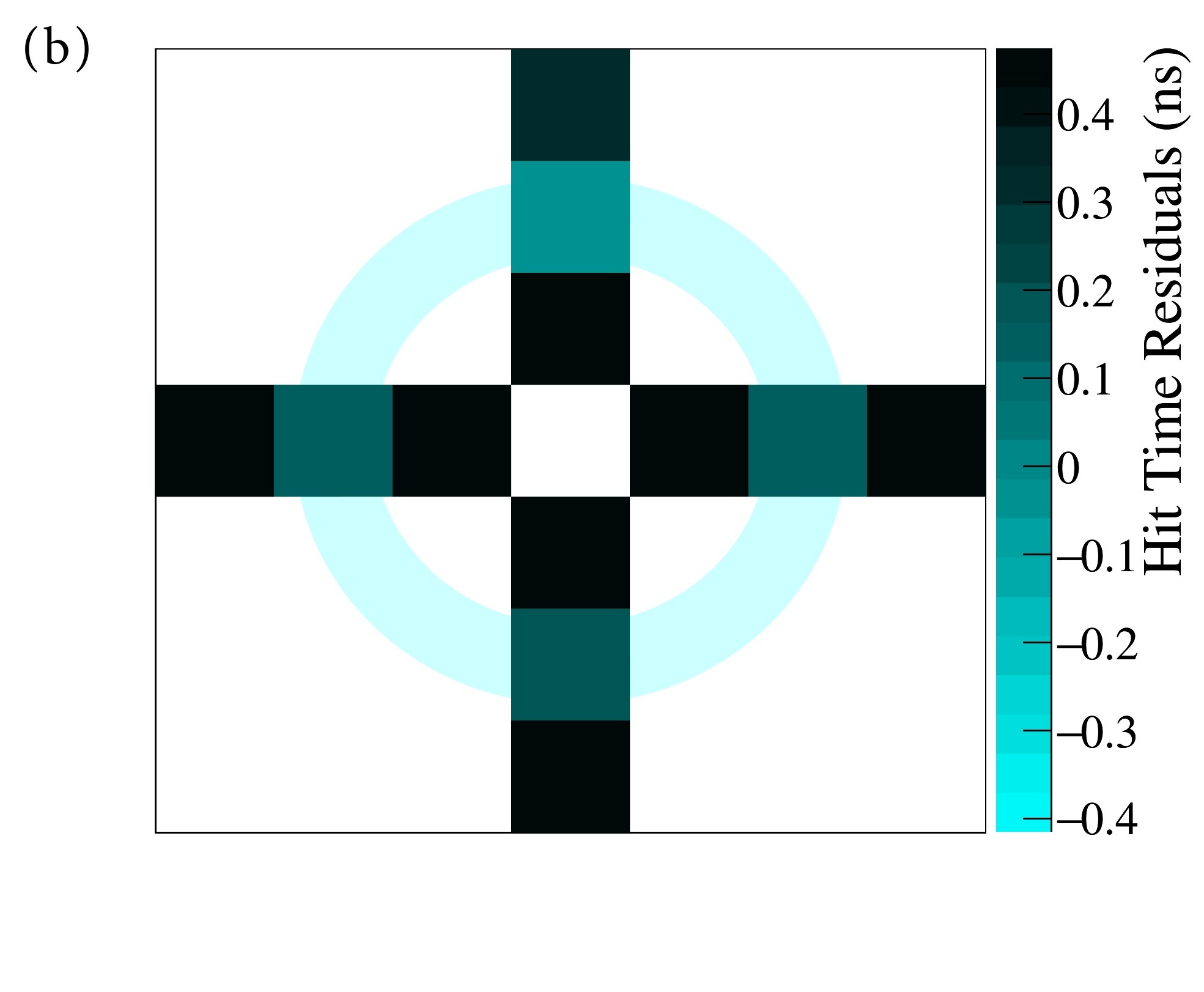}
	\caption{(a) Averaged number of detected PE and (b) averaged first-photon hit time residual per individual PMT for ring candidates in water data.}
	\label{fig:cosmics_water_npes}
\end{figure}

After application of the event selection criteria described in Sec.~\ref{s:event}, 137 ring candidates were selected in the water dataset.  The number of detected PEs and first photon hit-time residuals for a typical event are shown in \Cref{fig:cosmics_water_ring_candidate}. The averages across the data set for both the number of detected PEs per PMT and the hit-time residuals are shown in \Cref{fig:cosmics_water_npes}; both show a clear ring structure.

\begin{figure}
	\centering
	\includegraphics[width=0.9\columnwidth]{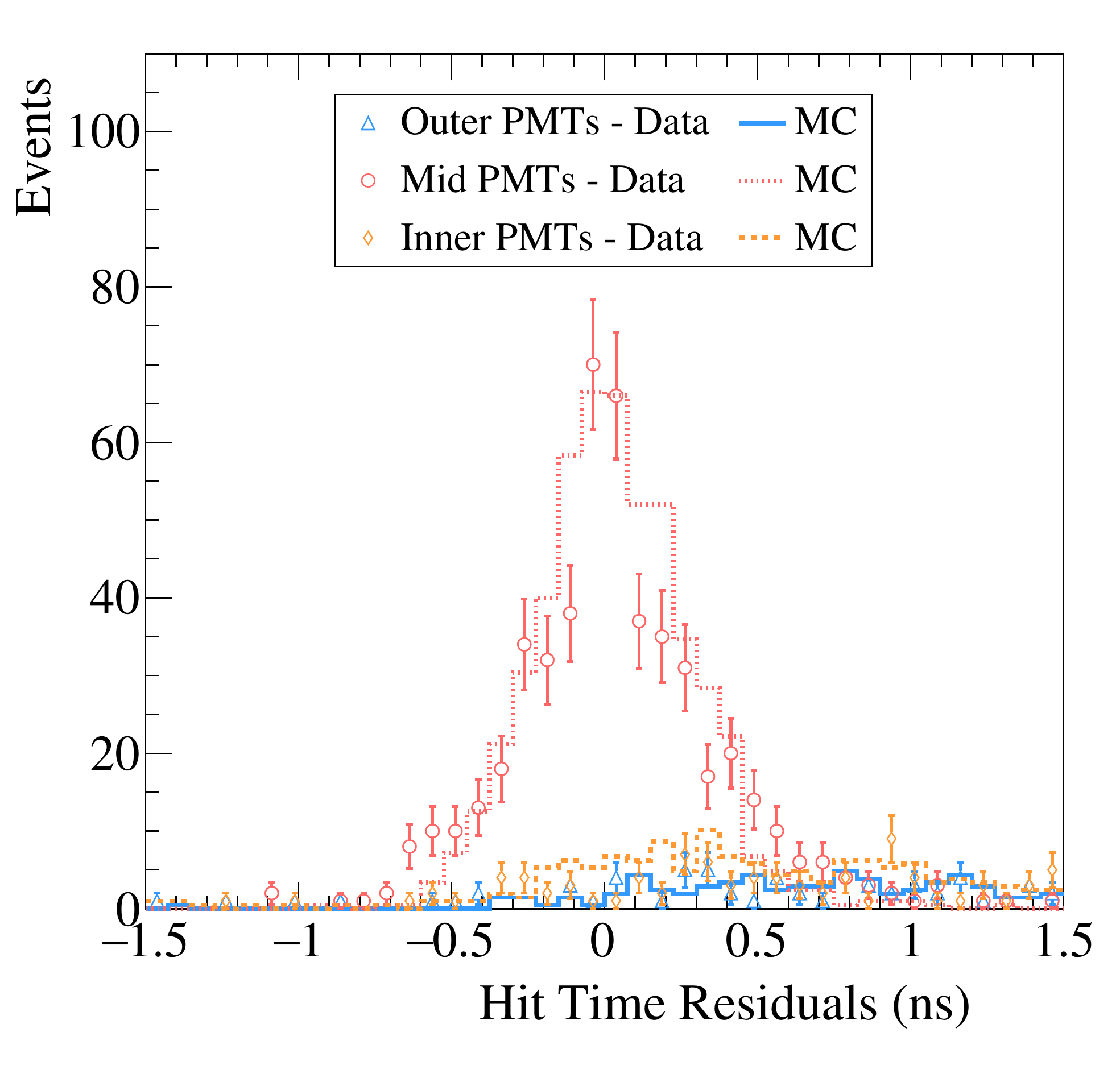}
	\caption{Distribution of hit time residuals for ring candidate events in water for each PMT grouping. Data points are shown with statistical errors, with the Monte Carlo prediction overlaid (dashed lines). }
	\label{fig:cosmics_water}
\end{figure}

The  distributions of hit-time residuals for each PMT radial group (inner, middle and outer PMTs) are shown in \Cref{fig:cosmics_water}, with the Monte Carlo prediction overlaid (with the additional smearing described in Sec.~\ref{s:recon}). 
The middle PMTs are the only ones detecting a sizable amount of light, and their time distribution is very sharp, compatible with the expected Cherenkov rings in water. The inner and outer PMTs are rarely hit.

The characteristic sharp time distribution of  Cherenkov light provides an excellent source for estimating the CHESS time resolution. The time distribution of the observed Cherenkov rings provides a measurement of $338\pm 12$~ps FWHM for the time precision of the CHESS detector. The limiting factor in this precision is the PMT TTS (300~ps FWHM). 

\section{Prospects with Liquid Scintillator Targets}\label{s:prospects}

An analysis of Monte Carlo data has been performed to predict the performance of CHESS for ring imaging and Cherenkov / scintillation separation with a pure LS target: using both LAB and LAB/PPO.
The targets are simulated using properties from~\cite{lab_emission,lab_timing,snop_private, labppo,labppo_quench}. The scintillation light yield is set to $1010$~photons/MeV for LAB~\cite{lab_timing} and $10800$~photons/MeV for LAB/PPO~\cite{snop_private}. 
Quenching is modeled using values for Birk's constant of $k_B = 0.0798$~mm/MeV in LAB and LAB/PPO~\cite{labppo_quench}. 
The emission spectra are shown in \Cref{fig:optics}. 
The time profile is modeled in the simulation as described in Sec.~\ref{sec:photon_prod}, with a rise time of $\tau_r = 1$~ns for LAB/PPO and and $\tau_r = 7.7$~ns for LAB, 
and decay times of $\tau_1 = 4.3$~ns, $\tau_2 = 16$~ns, $\tau_3 = 166$~ns \cite{labppo} for LAB/PPO, and $\tau_1 = 36.6$~ns, $\tau_2 = \tau_3 = 0$~\cite{lab_timing} for LAB.  

\begin{figure}
\centering
\includegraphics[width=\columnwidth]{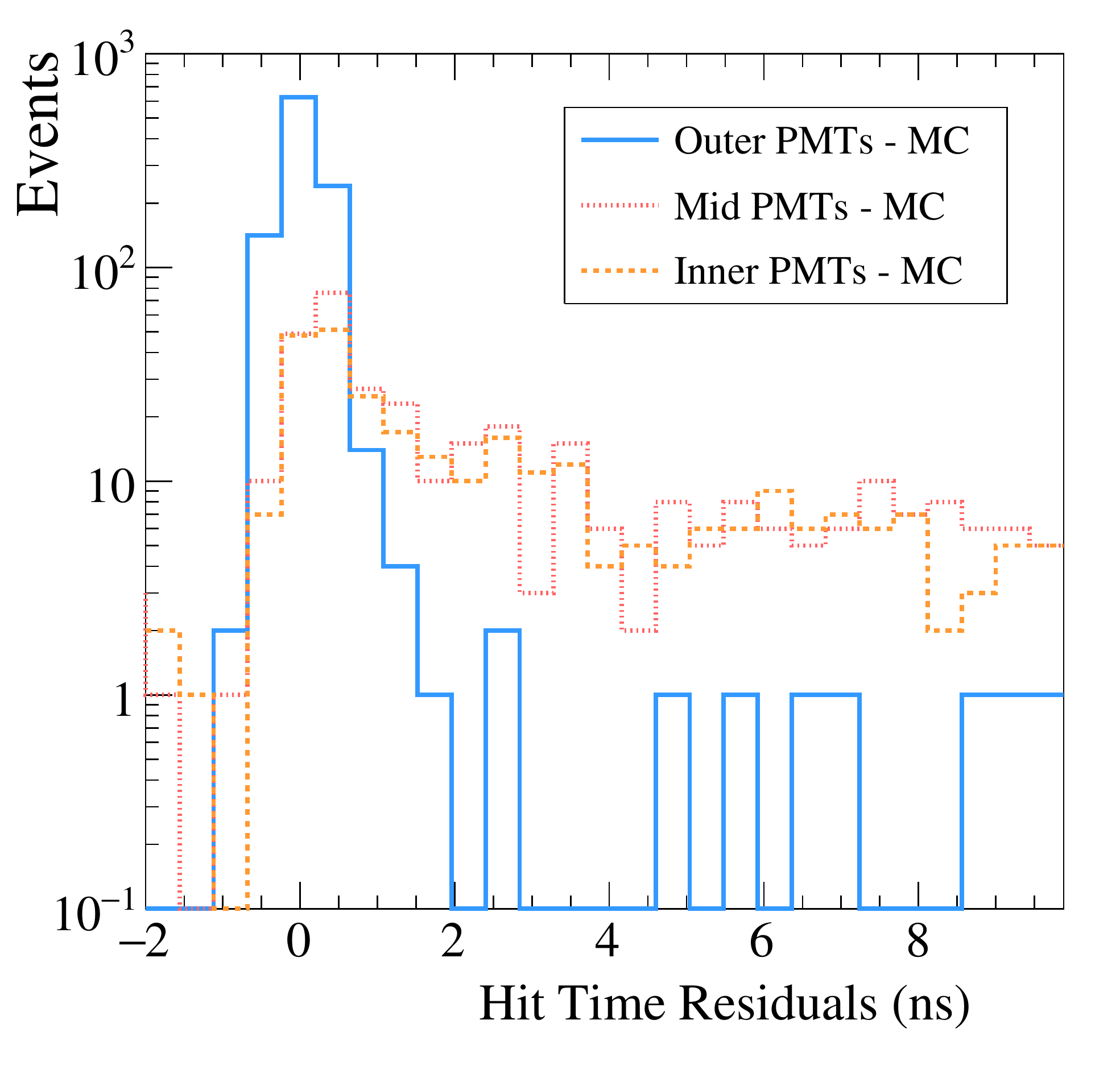}
\caption{Projected hit-time residual distributions for LAB as predicted from the Monte Carlo simulation.}
\label{f:LABtiming}
\end{figure}

\begin{figure}[!t]
\centering
\includegraphics[width=\columnwidth]{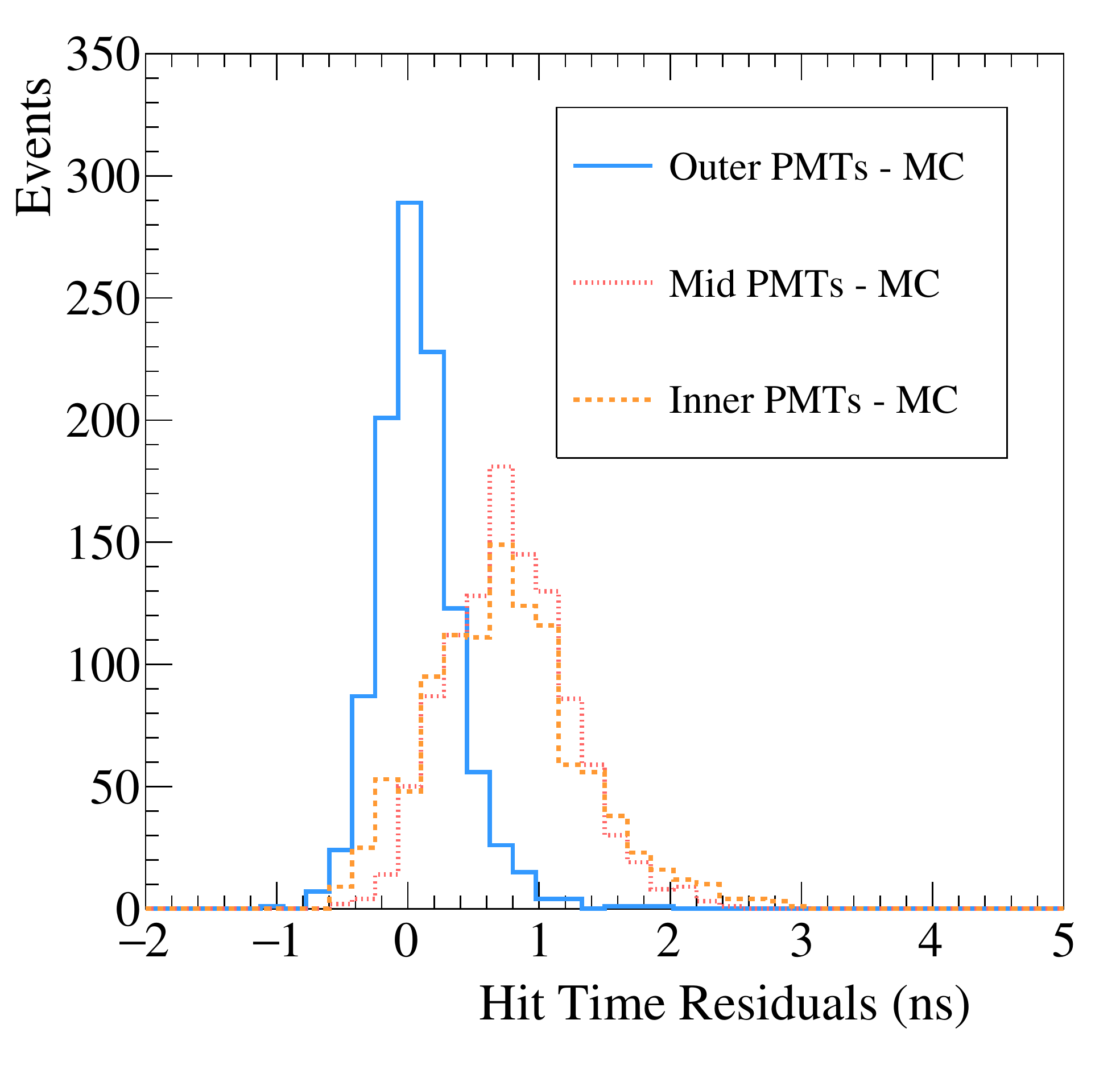}
\caption{Projected hit-time residual distributions for LAB/PPO as predicted from the Monte Carlo simulation.}
\label{f:LABPPOtiming}
\end{figure}

Several thousand cosmic muon events are simulated for each target, producing the hit-time residual distributions shown in Fig.~\ref{f:LABtiming} for LAB and Fig.~\ref{f:LABPPOtiming} for LAB/PPO. 
The small peak on the inner and middle PMTs seen in LAB is due to Cherenkov light contamination from follower events. 
The earliest hits are registered in the outer PMTs, where the Cherenkov ring is expected, while later features are due to scintillation light. 
Between 5 and 10 PE are expected on each PMT within the Cherenkov ring due to Cherenkov light alone.
The PMT hit-time residual is based on the first photon hit time for each channel,  thus with high confidence the hit-time of each PMT within the Cherenkov ring can be assigned to Cherenkov photon hits. Hits outside the Cherenkov ring are due to scintillation light.  
The separation of Cherenkov and scintillation photons can thus be defined by comparing the distributions of hit-time residuals on PMTs within the expected Cherenkov ring location (the outer PMTs for LAB and LAB/PPO) and those outside the ring (inner and middle PMTs).

 A timing cut can be developed to optimize this separation, selecting hits that occur before a time threshold, $t_c$.  The efficiency of identifying Cherenkov hits is defined as the fraction of outer PMT hits (Cherenkov hits) that occur before $t_c$.   The contamination due to scintillation photons  is evaluated as the fraction of all PMT hits  occuring before $t_c$ that are due to scintillation photons {\it i.e.} the number of inner and middle PMT hits occurring before $t_c$ divided by the total number of hits before $t_c$.

A value of $t_c=0.4$~ns results in a Cherenkov efficiency of $94\pm1\%$ in LAB and $81\pm1\%$ in LAB/PPO, with contaminations of $12\pm1$\% and $26\pm1$\%, respectively.  

A study has been performed in~\cite{asdc} of the impact of such separation on a potential large-scale NLDBD search.  This paper finds that solar neutrinos become a limiting background in a next-generation detector.  This dominant background can be rejected by use of the directional Cherenkov component, if separation of Cherenkov and scintillation light can be achieved.  Assuming a conservative rejection factor of two for these events, the study shows that sensitivity can be achieved down to the bottom of the inverted hierarchy (15~meV).  This study assumed a conservatively low light yield, based on the assumption that a WbLS target would be required in order to achieve the Cherenkov separation.  However, the work in this paper demonstrates the potential to identify the Cherenkov component even in a pure LS target, thus allowing directional sensitivity in a high light yield detector.  Such an experiment could have increased sensitivity to NLDBD, even into the normal hierarchy~\cite{steve}.
Ref.~\cite{chic} discusses an idea for how to use this separation to extract particle direction in a scintillator detector.

\section{Conclusions}\label{s:conc}

The time resolution of the CHESS experiment has been demonstrated to achieve the sub-ns precision required for successful separation of Cherenkov and scintillation light. The principle of Cherenkov ring imaging has been demonstrated using a deionized water target. A time resolution of $338\pm12$~ps FWHM has been achieved and clear Cherenkov rings have been detected in both charge and time on an event-by-event basis. A detailed simulation with LAB and LAB/PPO  shows that, with this time resolution,  time-based separation of the Cherenkov component from the scintillation light is possible with an efficiency of  $94\%$ and $81\%$, respectively, with scintillation contamination of 12\% and 26\%. 

If this separation can indeed be demonstrated in the experimental data then it would have significant implications for next-generation neutrino experiments.  Successful Cherenkov / scintillation separation would benefit a broad physics program, including: low-energy physics such as NLDBD and solar neutrinos; astrophysics topics such as supernova neutrinos and the diffuse supernova neutrino background; and high-energy physics such as nucleon decay, neutrino mass hierarchy, and CP violation.   Understanding the degree of separation that can be achieved, and quantifying how this varies with the specifics of the target cocktail will be critical steps in developing the program for a future large-scale experiment  such as \textsc{Theia}~\cite{theia}.

The next phase of CHESS will deploy pure LS in the target, followed by WbLS samples with varying LS fractions.  This will enable a full understanding of how the signal separation varies with cocktail.  In a future phase CHESS will be upgraded by populating the array with 12 additional fast PMTs. 
CHESS can also be used as a test bench for  studies of  next-generation MCP-based photon detectors~\cite{mcp, lappd, lappd2}.  The narrow pulse width of these detectors (10s of ps, compared to 5--10~ns for a typical PMT) could substantially improve the potential for charge-based separation by allowing much higher precision PE counting.

\section*{Acknowledgments}

This work was supported by the Laboratory Directed Research and Development Program of Lawrence Berkeley National Laboratory under U.S. Department of Energy Contract No. DEAC02- 05CH11231. 
The work conducted at Brookhaven National Laboratory was supported by the U.S. Department of Energy under contract DE-AC02-98CH10886. 
The authors would like to thank the SNO+ collaboration for providing data on the optical properties of LAB/PPO, including the light yield, absorption and reemission spectra, and refractive index.


\begin{thebibliography}{99}

\bibitem{imb} C.~Bratton {\it et al.}, Phys. Rev. D {\bf 37}, 3361 (1988).

\bibitem{superk} Y.~Fukuda {\it et al.} (Super-Kamiokande Collaboration), Phys. Rev. Lett. {\bf 81}, 1562 (1998).

\bibitem{sno} Q.~R.~Ahmad {\it et al.} (SNO Collaboration), Phys. Rev. Lett. {\bf 89}, 011301 (2002).

\bibitem{kamland} S.~Abe {\it et al.} (KamLAND Collaboration), Phys. Rev. Lett. {\bf 100}, 221803 (2008).

\bibitem{borexino} Borexino Collaboration, Nature {\bf 512}, 383 (2014).

\bibitem{lsnd} A.~Aguilar-Arevalo {\it et al.} (LSND Collaboration), Phys. Rev. D {\bf 64}, 112007 (2001).

\bibitem{cherenkov} P.~A.~\v{C}erenkov, Phys. Rev. {\bf 52}, 378 (1937).

\bibitem{birks} J.~B.~Birks, D.~Fry, L.~Costrell, and K.~Kandiah, {\it The Theory and Practice of Scintillation Counting} (Pergamon, 1964).

\bibitem{snop} S.~Andringa {\it et al.} (SNO+ Collaboration), Adv. High Energy Phys. {\bf 2016}, 6194250 (2016).

\bibitem{wbls} M.~Yeh, S.~Hans, W.~Beriguete, R.~Rosero, L.~Hu, R.~Hahn, M.~Diwan, D.~Jae, S.~Kettell, and L.~Littenberg, Nucl. Instrum. Methods {\bf A660}, 51 (2011).

\bibitem{asdc} J.~R. Alonso {\it et al.}, arXiv:1409.5864v3 [physics.ins-det] (2014).

\bibitem{theia} G. D. Orebi Gann for the THEIA interest group, in Proceedings of the Prospects in Neutrino Physics, arXiv:1504.08284 [physics.ins-det] (2015).

\bibitem{h11934} \url{https://www.hamamatsu.com/resources/pdf/etd/R11265U_H11934_TPMH1336E.pdf}.

\bibitem{lab_emission} C. Buck \& M.~Yeh, J. Phys. G: Nucl. Part. Phys. {\bf 43} 093001 (2016).

\bibitem{snop_private} SNO+ Collaboration (private communications).

\bibitem{mcp} B.~Adams, A.~Elagin, H.~Frisch, R.~Obaid, E.~Oberla, A.~Vostrikov, R.~Wagner, and M.~Wetstein, Nucl. Instum. Methods {\bf A732}, 392 (2013).

\bibitem{lappd} B.~W.~Adams {\it et al.} (LAPPD Collaboration), arXiv:1603.01843 [physics.ins-det] (2016), [Submitted to: JINST].

\bibitem{lappd2} O. H. W. Siegmund, J. B. McPhate, J. V. Vallerga, A. S. Tremsin, H. E. Frisch, J. W. Elam, A. U. Mane, and R. G. Wagner, J. Instrum. {\bf 9}, C04002 (2014).

\bibitem{cog-oring} \url{http://www.cog.de/fileadmin/downloads/COG_Resist_Folder_EN_Einzelseiten.pdf}.

\bibitem{ej550} \url{http://www.eljentechnology.com/index.php/products/accessories/ej-550-ej-552}.

\bibitem{h3164} \url{http://www.hamamatsu.com/us/en/product/alpha/P/3002/H3164-10/index.html}.

\bibitem{ej200} \url{http://www.eljentechnology.com/index.php/products/plastic-scintillators/ej-200-ej-204-ej-208-ej-212}.

\bibitem{9102ksb} \url{http://www.et-enterprises.com/photomultipliers}.

\bibitem{finemet} \url{http://www.hitachimetals.com/materials-products/amorphous-nanocrystalline/magnetic-shielding-sheet/documents/fm_shield_ms-f_ms-fr.pdf}.

\bibitem{buttonsource} \url{http://www.spectrumtechniques.com/products/sources/strontium-90/}.

\bibitem{wblsdaq}  B.~J.~Land, \url{http://github.com/BenLand100/WbLSdaq}, (2016).

\bibitem{caen-vme}  \url{http://www.caen.it/csite/CaenProd.jsp?idmod=689&parent=38}.

\bibitem{v1718}  \url{http://www.caen.it/csite/CaenProd.jsp?idmod=417&parent=11}.

\bibitem{hdf5}  The HDF Group, \url{http://www.hdfgroup.org} (1997-2016).

\bibitem{v6533}  \url{http://www.caen.it/csite/CaenProd.jsp?idmod=657&parent=23}.

\bibitem{v1730}  \url{http://www.caen.it/csite/CaenProd.jsp?parent=11&idmod=779}.

\bibitem{v1742}  \url{http://www.caen.it/csite/CaenProd.jsp?idmod=661&parent=11}.

\bibitem{drs4} \url{https://www.psi.ch/drs/}.

\bibitem{lecroy606zi} \url{http://teledynelecroy.com/oscilloscope/oscilloscopemodel.aspx?modelid=4782&capid=102&mid=504}.

\bibitem{ratpac} S. Seibert {\it et al.}, \url{http://rat.readthedocs.io/en/latest/}.

\bibitem{geant4} S. Agostinelli {\it et al.}, Nucl. Instrum. Methods Phys. Res. {\bf A506}, 250 (2003).

\bibitem{gaisser-mod} A. Tang, G. Horton-Smith, V. A. Kudryavtsev \& A. Tonazzo, Phys. Rev. D {\bf 74}, 053007 (2006);
 M. Guan {\it et al.}, arXiv:1509.06176 [hep-ex] (2015); G. Mengyun, C. Jun, Y. Changgen, S. Yaxuan \& K.-B. Luk, LBNL-4262E report.

\bibitem{glg4sim} G. Horton-Smith, \url{http://neutrino.phys.ksu.edu/~GLG4sim/}.

\bibitem{mcguire_palmer} R.~L.~McGuire and R.~C.~Palmer, IEEE Trans. Nucl. Sci. {\bf 14}, 217 (1967).

\bibitem{labppo} H.~M.~O'Keeffe, E.~O'Sullivan, and M.~C.~Chen, Nucl. Instum. Methods {\bf A640}, 119 (2011).

\bibitem{hamamatsu} Hamamatsu (private communication).

\bibitem{lab_timing} M.~Li {\it et al.}, arXiv:1511.09339 [physics.ins-det] (2015).

\bibitem{labppo_quench} B.~von~Krosigk, L.~Neumann, R.~Nolte, S.~R\"{o}ttger and K. Zuber, Eur. Phys. J. C {\bf 73}, 2390 (2013).

\bibitem{steve} S. D. Biller, Phys. Rev. D {\bf 87}, 071301 (2013).

\bibitem{chic} C. Aberle, A. Elagin, H. J. Frisch, M. Wetstein, \& L. Winslow, JINST \bf{9} (2014) P06012.

\end{thebibliography}
\end{document}